\documentclass[a4paper,12pt,twoside]{article}

\usepackage[british]{babel}
\usepackage{amssymb,amsmath,amsfonts}
\usepackage{graphicx}
\usepackage{subfigure}
\usepackage{enumerate}

\newcommand{\pluseq}{\hookleftarrow}
\def\vet#1{{\underline #1}}
\def\build#1_#2^#3{\mathrel{\mathop{\kern 0pt#1}\limits_{#2}^{#3}}}

\def\reali{\mathbb{R}}

\def\naturali{\mathbb{N}}
\def\interi{\mathbb{Z}}
\def\toro{\mathbb{T}}

\def\Ascr{\mathcal{A}}
\def\Bscr{\mathcal{B}}
\def\Cscr{\mathcal{C}}
\def\Dscr{\mathcal{D}}
\def\Fscr{\mathcal{F}}
\def\Gscr{\mathcal{G}}
\def\Hscr{\mathcal{H}}
\def\Jscr{\mathcal{J}}
\def\Kscr{\mathcal{K}}
\def\Lscr{\mathcal{L}}
\def\Oscr{\mathcal{O}}
\def\Pscr{\mathcal{P}}
\def\Rscr{\mathcal{R}}
\def\Tscr{\mathcal{T}}
\def\Uscr{\mathcal{U}}

\def\epsilon{\varepsilon}
\def\rho{\varrho}
\def\phi{\varphi}

\def\rgot{{\mathfrak r}}
\def\Bgot{{\mathfrak B}}
\def\Hgot{{\mathfrak H}}
\def\Kgot{{\mathfrak K}}

\def\Tgot{{\mathfrak T}}
\def\Xgot{{\mathfrak X}}

\def\lie#1{\Lscr_{#1}}

\def\fastpoisson#1#2{\left\{ #1,#2 \right\}_{\vet{L},\vet{\lambda}}}
\def\secpoisson#1#2{\left\{ #1,#2 \right\}_{\vet{\xi},\vet{\eta}}}

\def\wideitem#1{\par\hangindent\itemindent
   \noindent\hbox to\parindent{\hfil{#1}\enspace}\ignorespaces}
\title{\bf Secular dynamics of a planar model of the
  Sun-Jupiter-Saturn-Uranus system;\\ effective stability into the
  light of\\ Kolmogorov and Nekhoroshev theories}

\author{
{\bf ANTONIO GIORGILLI}\\
{\small Dipartimento di Matematica, Universit\`a degli Studi di Milano,}\\
{\small via Saldini 50, 20133\ ---\ Milano, Italy.}\\
{\bf UGO LOCATELLI}\\
{\small Dipartimento di Matematica, 
Universit\`a degli Studi di Roma ``Tor Vergata'',}\\
{\small via della Ricerca Scientifica 1, 00133\ ---\ Roma (Italy).}\\
{\bf MARCO SANSOTTERA}\\
{\small Dipartimento di Matematica, Universit\`a degli Studi di Milano,}\\
{\small via Saldini 50, 20133\ ---\ Milano, Italy.}\\
{\small e-mails:
  {\tt antonio.giorgilli@unimi.it, locatell@mat.uniroma2.it,}}\\
{\small {\tt marco.sansottera@unimi.it}}
}

\date{}

\begin{document}
\maketitle

\begin{abstract}
We investigate the long-time stability of the
Sun-Jupiter-Saturn-Uranus system by considering a planar secular
model, that can be regarded as a major refinement of the approach first
introduced by Lagrange.  Indeed, concerning the planetary orbital
revolutions, we improve the classical circular approximation by
replacing it with a solution that is invariant up to order two in the
masses; therefore, we investigate the stability of the secular system
for rather small values of the eccentricities. First, we explicitly
construct a Kolmogorov normal form, so as to find an invariant KAM
torus which approximates very well the secular orbits. Finally, we
adapt the approach that is at basis of the analytic part of the
Nekhoroshev's theorem, so as to show that there is a neighborhood of that
torus for which the estimated stability time is larger than the
lifetime of the Solar System. The size of such a neighborhood,
compared with the uncertainties of the astronomical observations,
is about ten times smaller.
\end{abstract}

\bigskip

\markboth{A. Giorgilli, U. Locatelli, M. Sansottera}{Secular${\scriptscriptstyle\ldots}$Sun-Jupiter-Saturn-Uranus system${\scriptscriptstyle\ldots}$Kolmogorov and Nekhoroshev theories}

\section{Introduction}\label{sec:intro}

The problem that we would like to investigate in this paper is
concerned with the long-time stability of the Solar System.  However
taking into account all planets (possibly including also the
satellites) is obviously an overwhelming task.  Therefore we limit our
efforts to considering a simplified model including three giant
planets, namely studying the Sun-Jupiter-Saturn-Uranus (SJSU) system.
Moreover we make a further simplification by considering the planar
model.  In the spirit of a long-time research that we are conducting,
we investigate the stability in Nekhoroshev sense in the neighbourhood
of an approximated KAM torus close to the real orbits of the planets.

To this end, we first make a short summary of the relevant
information concerning KAM and Nekhoroshev theories.  Then we give an
account of our results.

\subsection{General framework}\label{sbs:intro1}

In his celebrated article appeared in 1954
(see~\cite{Kolmogorov-1954}), Kolmogorov proved the persistence of
quasi-periodic motions on invariant tori for nearly-integrable
Hamiltonian systems.  In the same paper he pointed out the possible
applications of his theorem to the problem of the stability of the Solar
System.  More generally, Celestial Mechanics appeared as a main field
of applications of Kolmogorov theory.  Nowadays, there is no doubt
that the suggestion of Kolmogorov was prophetic.

Perhaps the first application is concerned with the stability of the
equilateral Lagrangian points in the planar circular restricted
three-body problem (hereafter PCR3BP, see~\cite{Leontovich-62}).  A
little later the papers of Moser and Arnold (see~\cite{Moser-1962}
and~\cite{Arnold-1963}) gave the first detailed proof of Kolmogorov's
theorem, that was only sketched in the original 1954 paper.  The three
above mentioned works marked the beginning of the so-called KAM theory.

The relevance of KAM theory for Celestial Mechanics was emphasized
also by Moser (see~\cite{Moser-1973}).  In particular he recalled the
old idea, strongly supported by Weierstrass, that the trigonometric
perturbation series of Celestial Mechanics should be actually
convergent, thus proving that the motion of the planets is
quasi-periodic.  That is, essentially described by the old-fashioned
epicycles of the Greek astronomy.

KAM theory has been used to prove the stability of some particular
systems: the spin-orbit problem, the motion of the asteroid Victoria
in the framework of the PCR3BP, the secular dynamics of the
Sun-Jupiter-Saturn system, etc. (see~\cite{Celletti-1994},
\cite{Cel-Chi-2007}, \cite{Loc-Gio-2000}, respectively).  All these
models have two degrees of freedom, therefore their stability is
proved by a topological argument. The key point is that an invariant
torus is a two-dimensional manifold that separates the
three-dimensional energy surface. Therefore, orbits starting in the
region between two invariant KAM tori on the same energy surface,
remain perpetually confined there.  For more degrees of freedom,
$n>2$, one is confronted with the problem that the set of invariant
KAM tori has positive Lebesgue measure, but empty interior. Therefore,
the energy manifold is ($2n-1$)-dimensional, while the invariant tori
are $n$-dimensional, so that they cannot act as barriers for the
motion.  This opens the problem of the so-called Arnold diffusion, the
generic existence of which has been recently proven near a resonance
of codimension one (see~\cite{Ber-Kal-Zha-2011} and references
therein).

A different approach to the problem of stability may be based on the
theorem of Nekhoroshev (see~\cite{Nekhoroshev-1977}
and~\cite{Nekhoroshev-1979}).  The difference may be summarized as
follows.  In KAM theory one proves perpetual stability only for a
large (Cantor-like) set of initial conditions, thus renouncing to
consider all possible configurations.  In Nekhoroshev theory one looks
for a result valid for all initial data (e.g., in an open set), but
accepts stability for a finite time, asking that time to be very long.
Actually, this kind of approach was already proposed by Moser and
Littlewood (see~\cite{Moser-1955}, \cite{Littlewood-1959.1}
and~\cite{Littlewood-1959.2}) in a local approach around an
equilibrium point.  Their theory essentially constitutes the so-called
analytic part of the proof of Nekhoroshev's theorem.  The remarkable
contribution of Nekhoroshev is represented by the geometric part,
which allows him to describe the dynamics in the whole phase space.

Let us rephrase the statement of Nekhoroshev's theorem in an informal
way.  The theorem is concerned with the general problem of dynamics,
so-named by Poincar\'e.  One considers an analytic Hamiltonian
$H(\vet{p},\vet{q})=h(\vet{p})+\epsilon f(\vet{p},\vet{q})$ (in
action-angle variables $(\vet{p},\vet{q})\in\Uscr\times\toro^n$, with
open $\Uscr\subset\reali^n$).  
  
\begin{quote}
  {\sl If $\epsilon$ is small enough and the unperturbed Hamiltonian
    $h(\vet{p})$ satisfies an appropriate steepness hypothesis, then
    for every orbit with initial condition
    $\big(\vet{p}(0),\vet{q}(0)\big)\in\Uscr\times\toro^n$ the bound
    $\|\vet{p}(t)-\vet{p}(0)\| < \epsilon^b$ holds true for $|t|\le
    T(\epsilon)$ where $T(\epsilon)\sim \exp(1/\epsilon^{a})$, for
    some constants $a, b \in (0,1)$.}
\end{quote}

Let us add some considerations concerning the application of the
concept of stability over finite, large time to physical systems.  In
this respect, the theory of Moser, Littlewood and Nekhoroshev can be
seen as an evolution of the old-fashioned adiabatic theory.  The
underlying idea is that the perpetual stability in Lyapunov sense is a
too strong property, which can hardly be proved.  In physical
applications, including a planetary system, it is enough to prove
stability for a time interval comparable with the lifetime of the
system itself.  E.g., for the Solar System the estimated age of the
Universe is enough.  This means that we should investigate stability
up to a time of about $10^{10}$ years.

Giving the adjective ``long'' a definite mathematical sense is of
course a more difficult task.  We can only rely on the dependence of
$T(\epsilon)$ on the perturbation parameter $\epsilon$.  With
reference to history we can collect a short list of attempts.

\begin{enumerate}[i.]
  \setlength{\itemsep}{0pt}
  \setlength{\parskip}{0pt}
  \setlength{\parsep}{0pt}
  
\item Adiabatic theory means $T(\epsilon)\sim 1/\epsilon$.  This is a
  concept that has been widely investigated in physics and played a
  relevant role in the development of Quantum Theory.

\item Birkhoff complete stability means $T(\epsilon)\sim 1/\epsilon^r$
  for some $r>1$.  The concept has been proposed by
  Birkhoff in~\cite{Birkhoff-1927}.\

\item Exponential stability means $T(\epsilon)\sim \exp(1/\epsilon^a)$
  as in the statement above.

\item Super-exponential stability means $T(\epsilon)\sim
  \exp\bigl(\exp(1/\epsilon^a)\bigr)$.  This has been proposed by
  Morbidelli and one of the authors in~\cite{Mor-Gio-1995}.  A further
  improvement by the same authors shows that in appropriate boxed
  subsets of the phase space one finds
  $T(\epsilon)\sim\exp\bigl(\exp\bigl(\ldots\exp(1/\epsilon^a)\bigr)\bigr)$ with
  an increasing number of exponentials.  The limit of the boxed
  subsets appears to be connected with the set of invariant KAM tori,
  see~\cite{Giorgilli-1997.4}.

\end{enumerate}

The possible applications to the real world deserve a careful detailed
discussion.  Unlike the mathematical approach, we must face the fact
that the size of the perturbation parameter is fixed by Nature.
Therefore the question may be formulated as follows: given that
$T(\epsilon)$ has some definite behaviour as $\epsilon$ goes to $0$, what can
we say for a specific system with a given value of $\epsilon$?  We refer
to this concept as {\it effective stability}, in the sense that we aim
to prove that $T(\epsilon)$, for that given value of $\epsilon$, is
large enough to cover some characteristic time of the physical system,
e.g., its lifetime, as we have already said.

The relevant fact in this connection is that the analytic form of
$T(\epsilon)$ provided by analytical theories appears to be just a
smoothing of a more complex behavior of the estimated stability time,
as it can be found with the help of computer algebra or similar
methods.  Let us explain this fact in the framework, e.g., of the
stability of an elliptic equilibrium.  The procedure goes through the
calculation of the Birkhoff normal form up to a finite order $r$.
Birkhoff's theory of complete stability states that in a neighbourhood
of radius $\rho>0$ of the equilibrium we get $T(\rho)\sim
1/(C_r\rho^r)$, with a constant $C_r$ that Birkhoff did not evaluate.
Here the natural perturbation parameter is the distance $\rho$ from
the equilibrium, which takes the place of $\epsilon$ in the general
statements above.  The crucial problem is the relation between $\rho$
and $r$.  What we can actually find is a function $\widetilde
T(\rho,r)$ depending on both parameters $\rho$ and $r$.  On the other
hand, having fixed $\rho$, we are allowed to make the best choice of
$r$ as a function of $\rho$ so as to maximize $\widetilde T(\rho,r)$.
In the case of the elliptic equilibrium one usually finds $C_r\sim
(r!)^c$ with $c\ge 1$, i.e., $\widetilde T(\rho,r)\sim
((r!)^c\rho^r)^{-1}$, and the optimal choice $r\sim (1/\rho)^{1/c}$
produces the exponential estimate $T(\rho)\sim\exp(1/\rho^a)$ with
$a=1/c$.  More precisely one finds that there is an increasing
sequence $\rho_1,\rho_2,\rho_3,\ldots$ of values of $\rho$ such that
in every interval $(\rho_j,\rho_{j+1})$ one gets $T(\rho)\sim
1/\rho^{\,j}$.  In this respect the theory of Moser, Littlewood and
Nekhoroshev appears to bound the latter function $T(\rho)$ from below.

In a practical application, if we are able to explicitly calculate the
Birkhoff normal form up to some maximal order $r$, e.g., using
computer algebra, then we can actually draw the function $T(\rho)$ as
the sequence of optimal values of $\widetilde T(\rho,r_j)$ in
different intervals, as we do later in Figure~\ref{fig:stab_time}.

Let us give a short historical account on the applications of the
methods above in Celestial Mechanics.  Most of them are concerned with
the dynamics of Trojan asteroids.  In~\cite{Giorgilli-1997}, a few
asteroids have been shown to be effectively stable (over the age of
the Universe) in the framework of the PCR3BP, where Sun and Jupiter
played the role of the primary bodies on circular orbits. Such an
approach is not limited to models having two degrees of freedom; in
fact, it has been extended to the spatial case
(see~\cite{Sko-Dok-2001}) and to the elliptic one
(see~\cite{Lho-Eft-Dvo-2008}, where the dynamics is represented by a
four-dimensional symplectic map instead of using a continuous
Hamiltonian flow). Let us recall that all these results are based on
the explicit construction of the Birkhoff normal form using computer
algebra.  By the way the case of the Lagrangian points is precisely
the one studied by Littlewood where an estimate similar to the
exponential one by Nekhoroshev has been found.  He commented:
\emph{``while not eternity, this is a considerable slice of it''}.

A similar approach allows us also to extend the theory of Lagrange and
Laplace for the secular motion of the longitudes of the perihelia and
nodes of the planets.  Indeed, following the traditional approach, we
may introduce the secular approximation by assuming that three
semi-major axes remain invariant up to order two in the masses.  Then
the Hamiltonian for the eccentricities and inclinations, with the
conjugate longitudes of the perihelia and nodes, may be written in
Poincar\'e variables as a system around an elliptic equilibrium
(see~\cite{Poincare-1905}).  The equilibrium in this case corresponds
to planar circular orbits.  Therefore we may investigate the effective
stability of the planets by just extending the method used for the
Lagrangian equilibria.  It should be noted that the complexity of the
problem becomes much larger in view of the increasing number of
degrees of freedom.  We stress however that the difficulty is a mere
technical one, due to the limitations of memory and computer power.

This method has been applied by the authors to a planar model of the
SJSU system (see~\cite{San-Loc-Gio-2013}).  However the results appear
neither realistic nor very promising due to the actual values of the
eccentricities of the planets. Indeed, in that paper we have shown
that the eccentricities of the planets are too large (about twice), in
order to ensure the effective stability of the secular dynamics over
the age of the Universe.  We remark that this is precisely the model
that we investigate in the present paper, with an additional
improvement that we are going to describe.

A productive combination of KAM and Nekhoroshev theories consists in
applying the usual, local theory for an elliptic equilibrium to the
neighbourhood of an invariant Kolmogorov torus.  In such a
neighbourhood, Kolmogorov procedure produces a Hamiltonian that may be
given the form $H(\vet{p},\vet{q})=\vet{\omega}\cdot\vet{p} +
h_1(\vet{p},\vet{q}) +h_2(\vet{p},\vet{q})+\ldots$, expanded in power
series of the actions $\vet{p}$ ($h_i$ being of degree $i+1$).  This
is indeed similar to the form of the Hamiltonian in a neighbourhood of
an elliptic equilibrium in action-angle variables.  The apparently
strong difference is that in the case of elliptic equilibrium we are
dealing with trigonometric polynomials $h_i$ of increasing orders,
while in the case of the torus $h_i$ is an infinite trigonometric
series.  However this represents a minor problem, indeed we can
suitably arrange the Hamiltonian as an expansion in trigonometric
polynomials, exploiting the exponential decay of Fourier
coefficients.  Therefore the analytic theory for the elliptic
equilibrium applies almost verbatim to the neighbourhood of a torus.
This remark suggests that the long-time stability of an elliptic
equilibrium and of a torus can be investigated using the same method.

We have applied the latter idea to the Sun-Jupiter-Saturn system
(see~\cite{Gio-Loc-San-2009}), where we used a previous result on the
existence of a torus of the SJS system (see~\cite{Loc-Gio-2007}) based
on the explicit expansion of the Hamiltonian and on the explicit
application of Kolmogorov method up to a finite, not too low order.
Then we worked out a Birkhoff normalization and showed that there is a
domain of effective stability, which is centered around an invariant
KAM torus.  The results were close to realistic ones.

In view of the previous experience, we decided to work out the application
to the planar SJSU system, exploiting the same idea of making
expansions around an approximated KAM torus.  This is indeed the goal
of the present paper.

\subsection{Plan of the work}\label{sbs:intro2}

As it is well known, the major problems in perturbation theory arise
from the existence of resonances in the trigonometric expansions.  It
goes without saying that a reliable theory should positively take into
account these resonant terms.  In the case of the SJSU system the main
resonances have been described by Murray and Holman. With the
numerical exploration made in~\cite{Mur-Hol-1999}, they have pointed out
the dynamical mechanism inducing a slightly chaotic component in the
motion of the major bodies of our planetary system. Actually, this
phenomenon is due to the overlap of some resonances involving three or
four bodies.  An example is given by the resonances
$$
3n_1-5n_2-7n_3+\left[(3-j)g_1+6g_2+jg_3\right]\,,
\qquad
{\rm with}\ j=0,\,1,\,2,\,3\ ,
$$ where $n_i$ stands for the mean motion frequency of the $i$-th
planet, $g_i$ means the (secular) frequency of its perihelion argument
and the labels $1,\,2,\,3$ refer to Jupiter, Saturn and Uranus,
respectively.  In fact, during the planetary motion each angle
corresponding to the resonances above jumps from libration to rotation
and vice versa.  Many other resonances analogous to the previous ones
are located in the vicinity of the real orbit of the system including
the Jovian planets, some of them involving also Neptune and the
frequencies related to the longitudes of the nodes.  In the same
article a rather simplistic argument is provided so as to evaluate the
time needed by these resonances to eject Uranus from the Solar System,
that is estimated to be about $10^{18}$ years. One should also recall
that the dynamics of the terrestrial planets is much more chaotic:
collisions between Mercury, Mars or, even, Venus with the Earth could
take place in about $3\,\times 10^9$ years (see~\cite{Las-Gas-2009}).

Our procedure is essentially based on two steps. First, we explicitly
perform the construction of the Kolmogorov normal form for the
planar secular model of the SJSU system.  In this first step, the
expansions of the Hamiltonians introduced by the normalization
algorithm are computed by using a software specially designed for
doing algebraic manipulations (see~\cite{Gio-San-Chronos-2012}). In the second
step, we avoid the explicit expansions (due to memory and power limits
of our computers) by setting up a suitable scheme of estimates for the
norms of the functions.  Precisely, we replace the explicit
construction of the Birkhoff normal form around the invariant KAM
torus with a recursive scheme of estimates on the norms.  The results
so produced are a little worse with respect to an explicit computation
of the series, but nevertheless the final results are acceptably close
to be realistic.  On the other hand associating to every function a
norm (i.e., a number) instead of a trigonometric polynomial obviously
makes the calculation definitely faster, while allowing to reach much
higher orders.

The whole procedure, including the optimization of the estimates with
an optimal choice of the expansion order $r$, allows us to evaluate a
lower bound for the stability time $T(\rho_0)$.  The final result is
the following.  We find a ball of effective stability over the age of
the Universe having radius $\rho_0$ and center on the previously found
KAM torus.  The value of $\rho_0$ is meaningful from a physical point
of view. Indeed, considering a ball of initial conditions that takes
into account the uncertainties of the astronomical observations, we
find a value of $\rho_0$ which is about ten times smaller.
Our result is not supported by a rigorous computer-assisted proof
(see, e.g.,~\cite{Cel-Chi-2007}), but we think that this could be done
with some additional effort, similarly to what we did in the
past (see~\cite{Loc-Gio-2000}).

The paper is organized as follows. In order to make the work rather
self-consistent, the model is introduced in section~\ref{sec:model},
where we put a particular care in pointing out some technical
difficulties.  Section~\ref{sec:kolmog} is devoted to the construction
of the Kolmogorov normal form, so as to find an invariant KAM torus
which approximates very well the secular orbits of our planetary
model.  In section~\ref{sec:nekh} we perform the search for stability
in the neighbourhood of the KAM torus.

\section{Settings for the definition of the Hamiltonian model}\label{sec:model}

For the sake of definiteness, in the present section we recall the
basic steps that are necessary to introduce the same planar secular
model of the Sun-Jupiter-Saturn-Uranus system that was already
studied in~\cite{San-Loc-Gio-2013}. We defer to sects.~2 and~3 of that
paper for more details.

\subsection{Classical expansion of the planar
  planetary Hamiltonian}\label{sbs:2D_plan_Ham}

Let us consider four point bodies $P_0,\,P_1,\,P_2,\,P_3$, with masses
$m_0,\,m_1,\,m_2,\,m_3$, mutually interacting according to Newton's
gravitational law.  Hereafter the indexes $0,\,1,\,2,\,3$ will
correspond to Sun, Jupiter, Saturn and Uranus, respectively. We
basically follow the formalism introduced by Poincar\'e (see,
e.g.,~\cite{Laskar-1989b} and~\cite{Las-Rob-1995} for a modern
exposition).  We remove the motion of the center of mass by using
heliocentric coordinates
$\vet{r}_j=\build{P_0P_j}_{}^{\longrightarrow}\,$, with
$j=1,\,2,\,3\,$.  Denoting by $\tilde{\vet{r}}_j$ the momenta
conjugate to $\vet{r}_j$, the Hamiltonian of the system has $6$
degrees of freedom, and reads
\begin{equation}
F(\tilde{\vet{r}},\vet{r})=
T^{(0)}(\tilde{\vet{r}})+U^{(0)}(\vet{r})+
T^{(1)}(\tilde{\vet{r}})+U^{(1)}(\vet{r}) \ ,
\label{Ham-iniz}
\end{equation}
where
$$
\begin{array}{rclrcl}
T^{(0)}(\tilde{\vet{r}}) &= &\frac{1}{2}\build{\sum}_{j=1}^{3} 
\frac{m_0+m_j}{m_0m_j}\,\|\tilde{\vet{r}}_j\|^2\ ,
\qquad
&T^{(1)}(\tilde{\vet{r}}) &=
&\frac{1}{m_0}\Big(\tilde{\vet{r}}_1\cdot\tilde{\vet{r}}_2+
\tilde{\vet{r}}_1\cdot\tilde{\vet{r}}_3+
\tilde{\vet{r}}_2\cdot\tilde{\vet{r}}_3\Big)\ ,
\cr\cr
U^{(0)}(\vet{r}) &= &-G\build{\sum}_{j=1}^{3}
\frac{m_0\, m_j}{\|\vet{r}_j\|}\ ,
\qquad
&U^{(1)}(\vet{r})
&= &-G\left(\frac{m_1\, m_2}{\|\vet{r}_1-\vet{r}_2\|}
+\frac{m_1\, m_3}{\|\vet{r}_1-\vet{r}_3\|}
+\frac{m_2\, m_3}{\|\vet{r}_2-\vet{r}_3\|}\right)\ .
\cr
\end{array}
$$

The plane set of Poincar\'e's canonical variables is defined as
\begin{equation}
\vcenter{\openup1\jot\halign{
 \hbox {\hfil $\displaystyle {#}$}
&\hbox {\hfil $\displaystyle {#}$\hfil}
&\hbox {$\displaystyle {#}$\hfil}
&\hbox to 6 ex{\hfil$\displaystyle {#}$\hfil}
&\hbox {\hfil $\displaystyle {#}$}
&\hbox {\hfil $\displaystyle {#}$\hfil}
&\hbox {$\displaystyle {#}$\hfil}\cr
\Lambda_j &=& \frac{m_0\, m_j}{m_0+m_j}\sqrt{G(m_0+m_j) a_j}\ ,
& &\lambda_j &=& M_j+\omega_j\ ,
\cr
\xi_j &=& \sqrt{2\Lambda_j}
\sqrt{1-\sqrt{1-e_j^2}}\,\cos\omega_j\ ,
& &\eta_j&=&-\sqrt{2\Lambda_j}
\sqrt{1-\sqrt{1-e_j^2}}\, \sin\omega_j\ ,
\cr
}}
\label{var-Poincare-piano}
\end{equation}
for $j=1\,,\,2\,,\,3\,$, where $a_j\,,\> e_j\,,\> M_j$ and $\omega_j$
are the semi-major axis, the eccentricity, the mean anomaly and the
perihelion longitude, respectively, of the $j$-th planet. Let us
remark that both $\xi_j$ and $\eta_j$ are of the same order of
magnitude as the eccentricity $e_j\,$.  Using Poincar\'e's
variables~(\ref{var-Poincare-piano}), the Hamiltonian $F$ can be
rearranged so that one has
\begin{equation}
F(\vet{\Lambda},\vet{\lambda},\vet{\xi},\vet{\eta})=
F^{(0)}(\vet{\Lambda})+
\mu F^{(1)}(\vet{\Lambda},\vet{\lambda},\vet{\xi},\vet{\eta}) \ ,
\label{Ham-iniz-Poincare-var}
\end{equation}
where $F^{(0)}=T^{(0)}+U^{(0)}$, $\mu F^{(1)}=T^{(1)}+U^{(1)}$.  Here,
the small dimensionless parameter
$\mu=\max\{m_1\,/\,m_0\,$, $\,m_2\,/\,m_0\,$, $\,m_3\,/\,m_0\,\}$ has been
introduced in order to highlight the different size of the terms
appearing in the Hamiltonian. According to the common language in
Celestial Mechanics, in the following we will refer to $\vet{\lambda}$
and to their conjugate actions $\vet{\Lambda}$ as the {\em fast
variables}, while $(\vet{\xi},\vet{\eta})$ will be called {\em
secular variables}.

We proceed now by expanding the
Hamiltonian~(\ref{Ham-iniz-Poincare-var}) in order to construct the
first basic approximation of Kolmogorov normal form.  We pick a
value $\vet{\Lambda}^*$ for the fast actions and perform a translation
$\Tscr_{\vet{\Lambda}^*}$ defined as
\begin{equation}
L_j=\Lambda_j-\Lambda_j^*\ ,
\qquad{\rm for}\ j=1\,,\, 2\,,\, 3\,.
\label{def-L}
\end{equation}
This is a canonical transformation that leaves the coordinates
$\vet{\lambda}\,$, $\vet{\xi}$ and $\vet{\eta}$ unchanged.  The
transformed Hamiltonian
$\Hscr^{(\Tscr)}=F\circ\Tscr_{\vet{\Lambda}^*}\,$ can be expanded in
power series of $\vet{L},\,\vet{\xi},\,\vet{\eta}$ around the origin.
Thus, forgetting an unessential constant we rearrange the Hamiltonian
of the system as
\begin{equation}
\Hscr^{(\Tscr)}(\vet{L},\vet{\lambda},\vet{\xi},\vet{\eta})=
\vet{n}^*\cdot\vet{L}+
\sum_{j_1=2}^{\infty}h_{j_1,0}^{({\rm Kep})}(\vet{L})+
\mu\sum_{j_1=0}^{\infty}\sum_{j_2=0}^{\infty}
h_{j_1,j_2}^{(\Tscr)}(\vet{L},\vet{\lambda},\vet{\xi},\vet{\eta}) \ ,
\label{Ham-trasl-fast}
\end{equation}
where the functions $h_{j_1,j_2}^{(\Tscr)}$ are homogeneous
polynomials of degree $j_1$ in the actions $\vet{L}$ and of degree
$j_2$ in the secular variables $(\vet{\xi},\vet{\eta})\,$. The
coefficients of such homogeneous polynomials do depend analytically
and periodically on the angles $\vet{\lambda}\,$.  The terms
$h_{j_1,0}^{({\rm Kep})}$ of the Keplerian part are homogeneous
polynomials of degree $j_1$ in the actions $\vet{L}\,$, the explicit
expression of which can be determined in a straightforward manner.  In
the latter equation the term which is both linear in the actions and
independent of all the other canonical variables (i.e.,
$\vet{n}^*\cdot\vet{L}$) has been separated in view of its relevance
in perturbation theory, as it will be discussed in the next
subsection.  We also expand the coefficients of the power series
$h_{j_1,j_2}^{(T_F)}$ in Fourier series of the angles
$\vet{\lambda}\,$, by following a traditional procedure in Celestial
Mechanics.  We work out these expansions for the case of the planar
SJSU system using a specially devised algebraic manipulation
(see~\cite{Gio-San-Chronos-2012} for an introduction to the main ideas
that have been translated in our codes).

\begin{table*}
\caption[]{Masses $m_j$ and initial conditions for Jupiter, Saturn and
  Uranus in our planar model.  We adopt the AU as unit of length, the
  year as time unit and set the gravitational constant
  $G=1\,$. With these units, the solar mass is equal to
  $(2\pi)^2$. The initial conditions are expressed by the usual
  heliocentric planar orbital elements: the semi-major axis $a_j\,$,
  the mean anomaly $M_j\,$, the eccentricity $e_j$ and the perihelion
  longitude $\omega_j\,$.  The data are obtained from those reported
  in Table~IV of~\cite{Standish-1998} by projecting them on the
  invariant plane (that is perpendicular to the total angular
  momentum) in a standard way.}
\label{tab:parameters_2D_SJSU}
\begin{center}
  \begin{tabular}{|c|l|l|l|}
\hline
& Jupiter ($j=1$) & Saturn ($j=2$) & Uranus ($j=3$)
\\
\hline
$m_{j}^{\phantom{\displaystyle 1}}$
& $(2\pi)^2/1047.355$
& $(2\pi)^2/3498.5$
& $(2\pi)^2/22902.98$
\\
$a_j$
& $5.20463727204700266$ & $9.54108529142232165$ & $19.2231635458410572$
\\
$M_j$
& $3.04525729444853654$ & $5.32199311882584869$ & $0.19431922829271914$
\\
$e_j$
& $0.04785365972484999$ & $0.05460848595674678$ & $0.04858667407651962$
\\
$\omega_j$
& $0.24927354029554571$ & $1.61225062288036902$ & $2.99374344439246487$ 
\\
\hline
\end{tabular}
\end{center}
\end{table*}

We now describe how to determine the fixed values $\vet{\Lambda}^*$
that allows us to perform the expansion~(\ref{Ham-trasl-fast}) of the
Hamiltonian as a function of the canonical coordinates
$(\vet{L},\vet{\lambda},\vet{\xi},\vet{\eta})$.  To this end we
perform a long-term numerical integration of Newton's equations
starting from the initial conditions related to the data reported in
Table~\ref{tab:parameters_2D_SJSU}.  After having computed the average
values $(a_1^*\,,\,a_2^*\,,\,a_3^*)$ of the semi-major axes during the
evolution, we determine the values $\vet{\Lambda}^*$ via the first
equation in~(\ref{var-Poincare-piano}).  In our calculations we
truncate the expansion as follows.  (a)~The Keplerian part is expanded
up to the quadratic terms.  The terms $h_{j_1,j_2}^{(\Tscr)}$ include:
(b1)~the linear terms in the actions $\vet{L}\,$, (b2)~all terms up to
degree~$18$ in the secular variables $(\vet{\xi},\vet{\eta})\,$,
(b3)~all terms up to the trigonometric degree $16$ with respect to the
angles $\vet{\lambda}\,$.  Our choice of the limits will be fully
motivated in the next subsection.

\subsection{The  secular model}\label{sbs:secular-model}
We now introduce the rather accurate description of the secular
dynamics provided by the average of the Hamiltonian up to order two in
the masses (see, e.g.,~\cite{Laskar-1988}, \cite{Laskar-1989c},
\cite{Loc-Gio-2000}, \cite{Kuz-Khol-2006}, \cite{Lib-Hen-2007}
and~\cite{Lib-San-2013}).  To this end we follow the approach
described in~\cite{Loc-Gio-2007}, carrying out two ``Kolmogorov-like''
normalization steps in order to eliminate the main perturbation terms
depending on the fast angles $\vet{\lambda}\,$.  We concentrate our
attention on the quasi-resonant angles $2\lambda_1-5\lambda_2\,$,
$\lambda_1-7\lambda_3$ and $3\lambda_1-5\lambda_2-7\lambda_3$, which
are the most relevant ones for the dynamics. The procedure is a little
cumbersome, and requires two main steps that we describe in the
following subsections.

\subsubsection{Partial reduction of the perturbation}\label{sss:Kolm-like-transf}
We emphasize that the Fourier expansion of the
Hamiltonian~(\ref{Ham-trasl-fast}) is generated just by terms due to
two-body interactions, and so harmonics including more than two fast
angles cannot appear.  Thus, at first order in the masses, only
harmonics with the quasi-resonant angles $2\lambda_1-5\lambda_2$ and
$\lambda_1-7\lambda_3$ do occur.  Actually, harmonics with the
quasi-resonant angle $3\lambda_1-5\lambda_2-7\lambda_3$ are generated
by the first Kolmogorov-like transformation, but are of second order
in the masses, and should be removed by the second Kolmogorov-like
transformation described in the next section.

Let us go into details.  We denote by $\big\lceil f
\big\rceil_{\vet{\lambda};K_{F}}$ the Fourier expansion of a function
$f$ truncated so as to include only its harmonics
$\vet{k}\cdot\vet{\lambda}$ satisfying the restriction $0<|\vet{k}|\le
K_{F}\,$, with some fixed $K_F$, being
$|\vet{k}|=|k_1|+|k_2|+|k_3|\,$.  We also denote by
$\langle\cdot\rangle_{\vet{\lambda}}$ the average with respect to the
angles $\lambda_1\,$, $\lambda_2$ and $\lambda_3\,$.  The canonical
transformations are using the Lie series algorithm (see,
e.g.,~\cite{Giorgilli-1995}).

We set $K_{F}=8$ and transform the Hamiltonian~(\ref{Ham-trasl-fast})
as $\hat \Hscr^{(\Oscr 2)}=\exp\lie{\mu\,\chi_{1}^{(\Oscr 2)}}\,
\Hscr^{(\Tscr)}$ with the generating function $\mu\,\chi_{1}^{(\Oscr
2)}(\vet{\lambda},\vet{\xi},\vet{\eta})$ determined by solving the
equation
\begin{equation}
\sum_{j=1}^{3}n^*_j\,
\frac{\partial\,\chi_{1}^{(\Oscr 2)}}{\partial \lambda_j}
+\sum_{j_2=0}^{6}\left\lceil h_{0,j_2}^{(\Tscr)}
\right\rceil_{\vet{\lambda};8}
(\vet{\lambda},\vet{\xi},\vet{\eta})=0\ .
\label{eqperchi1Oscr2}
\end{equation}
Notice that, by definition, $\big\langle\big\lceil f
\big\rceil_{\vet{\lambda};K_{F}}\big\rangle_{\vet{\lambda}}=0\,$,
which assures that equation~(\ref{eqperchi1Oscr2}) can be solved
provided the frequencies $n_1^*\,,\,n_{2}^*$ and $n_{3}^*$ are
non-resonant up to order $8\,$, as it actually occurs in our planar
model of the SJSU system. The Hamiltonian $\hat \Hscr^{(\Oscr 2)}$ has
the same form of $\Hscr^{(\Tscr)}$ in~(\ref{Ham-trasl-fast}), with the
functions $h_{j_1,j_2}^{(\Tscr)}$ replaced by new ones, that we denote
by $\hat{h}_{j_1,j_2}^{(\Oscr 2)}$, generated by expanding the Lie
series $\exp\lie{\mu\,\chi_{1}^{(\Oscr 2)}}\,\Hscr^{(\Tscr)}$ and by
gathering all the terms having the same degree both in the fast
actions and in the secular variables.

Now we perform a second canonical transformation $\Hscr^{(\Oscr
2)}=\exp\lie{\mu\,\chi_{2}^{(\Oscr 2)}}\,\hat \Hscr^{(\Oscr 2)}\,$,
where the generating function $\mu\,\chi_{2}^{(\Oscr
2)}(\vet{L},\vet{\lambda},\vet{\xi},\vet{\eta})$ (which is linear
with respect to $\vet{L}$) is determined by solving the equation
\begin{equation}
\sum_{j=1}^{3}n^*_j\,
\frac{\partial\,\chi_{2}^{(\Oscr 2)}}{\partial \lambda_j}
+\sum_{j_2=0}^{6}\left\lceil \hat{h}_{1,j_2}^{(\Oscr 2)}
\right\rceil_{\vet{\lambda};8}
(\vet{L},\vet{\lambda},\vet{\xi},\vet{\eta})=0\ .
\label{eqperchi2Oscr2}
\end{equation}
Again, the Hamiltonian $\Hscr^{(\Oscr 2)}$ can be written in a form
similar to~(\ref{Ham-trasl-fast}), namely
\begin{equation}
\Hscr^{(\Oscr 2)}(\vet{L},\vet{\lambda},\vet{\xi},\vet{\eta})=
\vet{n}^*\cdot\vet{L}+
\sum_{j_1=2}^{\infty} h_{j_1,0}^{({\rm Kep})}(\vet{L})+
\mu\sum_{j_1=0}^{\infty}\sum_{j_2=0}^{\infty}
h_{j_1,j_2}^{(\Oscr 2)}
(\vet{L},\vet{\lambda},\vet{\xi},\vet{\eta};\mu)\ .
\label{Ham-Omu^2}
\end{equation}
where the new functions $h_{j_1,j_2}^{(\Oscr 2)}$ are calculated as
previously explained for $\hat{h}_{j_1,j_2}^{(\Oscr 2)}\,$.  Moreover,
they still have the same dependence on their arguments as
$h_{j_1,j_2}^{(\Tscr)}$ in~(\ref{Ham-trasl-fast}).

If terms of second order in $\mu$ are neglected, then the Hamiltonian
$\Hscr^{(\Oscr 2)}$ possesses the secular three-dimensional invariant
torus $\vet{L}=\vet{0}$ and $\vet{\xi}=\vet{\eta}=\vet{0}$.  Thus, in
a small neighborhood of the origin of the translated fast actions and
for small eccentricities the solutions of the system with Hamiltonian
$\Hscr^{(\Oscr 2)}$ differ from those of its average $\langle
\Hscr^{(\Oscr 2)}\rangle_{\vet{\lambda}}$ by a quantity
$\Oscr(\mu^2)$.  In this sense the average of the
Hamiltonian~(\ref{Ham-Omu^2}) approximates the real dynamics of the
secular variables up to order two in the masses, and due to the choice
$K_{F}=8$ takes into account the quasi-resonances $5:2$ between
Jupiter and Saturn and $7:1$ between Jupiter and Uranus.

In this part of the calculation we produce a truncated series which is
represented as a sum of monomials
$$
c_{\vet{j},\vet{k},\vet{r},\vet{s}}\,L_1^{j_1} L_2^{j_2} L_3^{j_3} 
 \,\xi_1^{r_1} \xi_2^{r_2} \xi_3^{r_3}
  \,\eta_1^{s_1} \eta_2^{s_2} \eta_3^{s_3}
   \,{\scriptstyle{{\displaystyle{\sin}}\atop{\displaystyle{\cos}}}}
     (k_1\lambda_1+k_2\lambda_2+k_3\lambda_3)\ .
$$ The truncated expansion of~$\Hscr^{(\Oscr 2)}$ contains
$94\,109\,751$ such monomials.  We truncate our expansion at degree 16
in the fast angles $\vet{\lambda}$ (keeping all harmonics) and at degree 18 in the slow
variables $\vet{\xi},\,\vet{\eta}$ (we shall justify this choice at
the end of the next section).

\subsubsection{Second approximation and reduction to the secular
  Hamiltonian}\label{sss:red-sec-Ham} Since we plan to consider the
secular system, we perform a partial average by keeping only the main
terms that contain the quasi-resonant angle
$3\lambda_1-5\lambda_2-7\lambda_3$.  More precisely, we first consider
the reduced Hamiltonian
\begin{equation}
\left\langle \Hscr^{(\Oscr 2)}\big|_{\vet{L}
 =\vet{0}}\,\right\rangle_{\vet{\lambda}}=
\mu\sum_{j_2=0}^{\infty} \big\langle
h_{0,j_2}^{(\Oscr 2)}(\vet{\xi},\vet{\eta};\mu)
 \big\rangle_{\vet{\lambda}}\ ,
\label{sec-Ham-Omu^2}
\end{equation}
namely we set $\vet{L}=\vet{0}$ and average $\Hscr^{(\Oscr 2)}$ by
removing all the Fourier harmonics depending on the angles.  Next, we
select in $\Hscr^{(\Oscr 2)}$ the Fourier harmonics that contain the
wanted quasi-resonant angle $3\lambda_1-5\lambda_2-7\lambda_3$ and add
them to the Hamiltonian~(\ref{sec-Ham-Omu^2}).  Finally, we perform on
the resulting Hamiltonian the second Kolmogorov-like step.  With more
detail, this is the procedure, which is an adaptation of a scheme
already used in~\cite{Loc-Gio-2000}.  For $(j_1,j_2)\in\naturali^2$ we
select the quasi-resonant terms
\begin{equation}
\vcenter{\openup1\jot\halign{
 \hbox {\hfil $\displaystyle {#}$}
&\hbox {\hfil $\displaystyle {#}$\hfil}
&\hbox {$\displaystyle {#}$\hfil}\cr
\mu^2 h_{j_1,j_2}^{({\rm q.r.})}
  (\vet{L},\vet{\lambda},\vet{\xi},\vet{\eta}) &=
&\mu\,\big\langle\, h_{j_1,j_2}^{(\Oscr 2)}
\,\exp\big[-{\rm i}(3\lambda_1-5\lambda_2-7\lambda_3)\big]
\,\big\rangle_{\vet{\lambda}}
\,\exp\big[{\rm i}(3\lambda_1-5\lambda_2-7\lambda_3)\big]\,+
\cr
& &\mu\,\big\langle\, h_{j_1,j_2}^{(\Oscr 2)}
\,\exp\big[{\rm i}(3\lambda_1-5\lambda_2-7\lambda_3)\big]
\,\big\rangle_{\vet{\lambda}}
\,\exp\big[-{\rm i}(3\lambda_1-5\lambda_2-7\lambda_3)\big]
\ .
\cr
}}
\label{def-filter-terms-Omu^2}
\end{equation}
Actually, this means that in our expression we just remove all
monomials but the ones containing the wanted quasi-resonant angle.
Using the selected terms we determine a generating function
$\mu^2\chi_{1}^{({\rm q.r.})}(\vet{\lambda},\vet{\xi},\vet{\eta})$ by
solving the equation
\begin{equation}
\sum_{j=1}^{3}n^*_j\,
\frac{\partial\,\chi_{1}^{({\rm q.r.})}}{\partial \lambda_j}
+\sum_{j_2=0}^{9} h_{0,j_2}^{({\rm q.r.})}(\vet{\lambda},\vet{\xi},\vet{\eta})=0\ .
\label{eqperchi1filt}
\end{equation}
Here we make the calculation faster by keeping only terms up to degree
$9$ in~$(\vet{\xi},\vet{\eta})\,$, this allows us to keep the more
relevant quasi-resonant contributions.  Then, we calculate only the
interesting part of the transformed Hamiltonian
$\exp\lie{\mu^2\,\chi_{1}^{({\rm q.r.})}}\,\Hscr^{(\Oscr 2)}\,$,
namely we keep in the transformation only the part which is
independent of all the fast variables $(\vet{L},\vet{\lambda})\,$.
This produces the secular Hamiltonian $\Hscr^{({\rm sec})}$,
independent of $\vet{\lambda}$, which
satisfies the formal equation $\big\langle
\exp\lie{\mu^2\,\chi_{1}^{({\rm q.r.})}}\,\Hscr^{(\Oscr
  2)}\big\rangle_{\vet{\lambda}}= \Hscr^{({\rm
    sec})}+\Oscr(\|\vet{L}\|)+o(\mu^4)\,$, where
\begin{equation}
\vcenter{\openup1\jot\halign{
 \hbox {\hfil $\displaystyle {#}$}
&\hbox {\hfil $\displaystyle {#}$\hfil}
&\hbox {$\displaystyle {#}$\hfil}\cr
\Hscr^{({\rm sec})}(\vet{\xi},\vet{\eta})
&= &\mu\sum_{j_2=0}^{\infty} \big\langle
h_{0,j_2}^{(\Oscr 2)}\big\rangle_{\vet{\lambda}}+
\mu^4\Bigg\langle\,\frac{1}{2}
\fastpoisson{\chi_{1}^{({\rm q.r.})}}
{\Lscr_{\mu^2\,\chi_{1}^{({\rm q.r.})}}h_{2,0}^{({\rm Kep})}}
+\cr
&
&\ \fastpoisson{\chi_{1}^{({\rm q.r.})}}{\sum_{j_2=0}^{\infty} h_{1,j_2}^{({\rm q.r.})}}
+\frac{1}{2}\secpoisson{\chi_{1}^{({\rm q.r.})}}
{\sum_{j_2=0}^{\infty} h_{0,j_2}^{({\rm q.r.})}}
\,\Bigg\rangle_{\vet{\lambda}}\ .
\cr
}}
\label{def-sec-Ham}
\end{equation}
Here, we denoted by $\fastpoisson{\cdot}{\cdot}$ and
$\secpoisson{\cdot}{\cdot}$ the terms of the Poisson bracket involving
only the derivatives with respect to the conjugate variables
$(\vet{L},\vet{\lambda})$ and $(\vet{\xi},\vet{\eta})$, respectively.

The Hamiltonian so constructed is the secular one, describing the slow
motion of eccentricities and perihelia.  In view of the d'Alembert rules
$\Hscr^{({\rm sec})}$ contains only terms of even degree and so the
lowest order significant term has degree 2. By using our specially
designed algebraic manipulator, we have computed the power series
expansion of the secular Hamiltonian up to degree 18 in the slow
variables.  In order to allow comparisons with other expansions, our
results up to degree~4 in~$(\vet{\xi},\vet{\eta})$ are reported in
appendix~A of~\cite{San-Loc-Gio-2013}.

Let us close this section with a few remarks which justify our choice
of the truncation orders.  The limits on the expansions in the fast
actions $\vet{L}$ have been illustrated at points~(a) and~(b1) at the
end of sect.~\ref{sbs:2D_plan_Ham}, and they are the smallest ones
that are required in order to make the Kolmogorov-like normalization
procedure significant.  Since we want to keep the quasi-resonant
angles $2\lambda_1-5\lambda_2\,$, $\lambda_1-7\lambda_3$ and
$3\lambda_1-5\lambda_5-7\lambda_3\,$, we should set the truncation
order for Fourier series to $16$, which is enough.  The choice to
truncate the polynomial expansion at degree $18$ in the secular
variables $(\vet{\xi},\vet{\eta})$ is somehow subtler.  In view of
d'Alembert rules the harmonics $2\lambda_1-5\lambda_2$ and
$\lambda_1-7\lambda_3$ have coefficients of degree at least 3 and 6,
respectively, in the secular variables.  Thus, we decided to calculate
the generating functions $\chi_{1}^{(\Oscr 2)}$ and $\chi_{2}^{(\Oscr
  2)}$ up to degree $6$ (recall equations~(\ref{eqperchi1Oscr2})
and~(\ref{eqperchi2Oscr2})). Furthermore, the quasi-resonant angle
$3\lambda_1-5\lambda_5-7\lambda_3$ does not appear initially in the
Hamiltonian, but is generated by Poisson bracket between the harmonics
$2\lambda_1-5\lambda_2$ and $\lambda_1-7\lambda_3$, which produces
monomials of degree $9$ in $(\vet{\xi},\vet{\eta})$.  Therefore, we
decided to calculate the generating function $\chi_{1}^{({\rm q.r.})}$ up
to degree $9$ (recall equation~(\ref{eqperchi1filt})).  Finally, in the second Kolmogorov-like
step we want to keep the secular terms generated by the harmonic
$3\lambda_1-5\lambda_5-7\lambda_3$, which are produced by Poisson
bracket between monomials containing precisely this harmonic, and then
the result has maximum degree $18$ in $(\vet{\xi},\vet{\eta})$.  This
justifies the final truncation order for the slow variables.

\section{Normalization algorithm constructing invariant tori close to
an elliptic equilibrium point}\label{sec:kolmog}
The lowest order approximation of the secular Hamiltonian
$\Hscr^{({\rm sec})}$, namely its quadratic term, is essentially the
one considered in the theory first developed by Lagrange
(see~\cite{Lagrange-1776}), later extended by Laplace
(see~\cite{Laplace-1772}, \cite{Laplace-1784} and~\cite{Laplace-1785})
and further improved by Lagrange himself (see~\cite{Lagrange-1781},
\cite{Lagrange-1782}).  In modern language, we say that the origin of
the reduced phase space (i.e.,
$(\vet{\xi},\vet{\eta})=(\vet{0},\vet{0})\,$) is an elliptic
equilibrium point (for a review using a modern formalism, see sect.~3
of~\cite{Bia-Chi-Val-2006}, where a planar model of our Solar System
is considered).

It is well known that (under mild assumptions on the quadratic part of
the Hamiltonian which are satisfied in our case) one can find a linear
canonical transformation
$(\vet{\xi},\vet{\eta})=\Dscr(\vet{x},\vet{y})$ which diagonalizes
the quadratic part of the Hamiltonian, so that we can write
$\Hscr^{({\rm sec})}$ in the new coordinates as
\begin{equation}
H^{(\Dscr)}(\vet{x},\vet{y})=\sum_{j=0}^{3}\frac{\nu_j}{2}\left(x_j^2+y_j^2\right)+
H_{2}^{(\Dscr)}(\vet{x},\vet{y})+H_{4}^{(\Dscr)}(\vet{x},\vet{y})+
H_{6}^{(\Dscr)}(\vet{x},\vet{y})+\ldots\ ,
\label{Ham^0}
\end{equation}
where $\nu_j$ are the secular frequencies in the small oscillations
limit and $H_{2s}^{(\Dscr)}$ is a homogeneous polynomial of degree $2s+2$
in $(\vet{x},\vet{y})\,$.  The calculated values of
$\nu_1,\nu_2$ and $\nu_3)$ are reported in
Table~\ref{tab:freq_condiniz_sec_2D_SJSU}.

\begin{table*}
\caption[]{Angular velocities $\vet{\nu}$ and initial conditions
  $(\vet{x}(0),\vet{y}(0))$ for our planar secular model about the
  motions of Jupiter, Saturn and Uranus. The frequency vector
  $\vet{\nu}$ refers to the harmonic oscillators approximation of
  the Hamiltonian $H^{(\Dscr)}$ (written in~(\ref{Ham^0})) and its values
  are given in rad/year.}
\label{tab:freq_condiniz_sec_2D_SJSU}
\begin{center}
\begin{tabular}{|c|l|l|l|}
\hline
& \hfil$j=1$ &\hfil$j=2$ & \hfil$j=3$
\\
\hline
$\nu_{j}^{\phantom{\displaystyle 1}}$
& $-1.1212724892\,\times 10^{-4}$
& $-1.9688444678\,\times 10^{-5}$
& $-1.1134564418\,\times 10^{-5}$
\\
$x_j(0)$
& $\phantom{-}1.5407573458\,\times 10^{-2}$
& $-3.0574059274\,\times 10^{-2}$
& $\phantom{-}1.1186486403\,\times 10^{-2}$
\\
$y_j(0)$
& $-2.5320810665\,\times 10^{-2}$
& $-5.2728862107\,\times 10^{-3}$
& $\phantom{-}6.0669645406\,\times 10^{-3}$
\\
\hline
\end{tabular}
\end{center}
\end{table*}

At this point, it is convenient to introduce a further preliminary
canonical transformation $(\vet{x},\vet{y})=\Ascr(\vet{I},\vet{\phi})$
so as to introduce action-angle coordinates, namely $x_j =
\sqrt{2I_j}\cos\phi_j$ and $y_j = \sqrt{2I_j}\sin\phi_j$, for
$j=1,\,2,\,3$. Then, the expansion of the new Hamiltonian
$H^{({\rm I})}=H^{(\Dscr)}\circ\Ascr$ is (sometimes said) of
d'Alembert type, i.e., it can be written as
\begin{equation}
H^{({\rm I})}(\vet{I},\vet{\phi}) = \vet{\nu}\cdot\vet{I} +
\sum_{s\ge 1} f_{2s}^{({\rm I})}(\vet{I},\vet{\phi})\ ,
\label{eq:ham_BNF_deg_2}
\end{equation}
where the functions $f_{2s}^{({\rm I})}$ are homogeneous polynomials
of degree $2s+2$ in the square root of the actions and trigonometric
polynomials of degree $2s+2$ in the angles. Moreover, for any fixed
index $m\in\{1,\,\ldots,\,n\}$ (obviously being, in our case, the
number of degrees of freedom $n=3$) and for all term appearing in the
expansion of $f_{2s}^{({\rm I})}$ the $m$-th component of the Fourier
harmonics is not greater than the corresponding degree in $\sqrt{I_m}$
and they have the same parity, i.e.,
\begin{equation}
\vcenter{\openup1\jot\halign{
 \hbox {\hfil $\displaystyle {#}$}
&\hbox {\hfil $\displaystyle {#}$\hfil}
&\hbox {$\displaystyle {#}$\hfil}\cr
f_{2s}^{({\rm I})}(\vet{I},\vet{\phi}) & = & 
\sum_{i_1+\ldots+i_n=2s+2}\,\sum_{j_1=0}^{i_1}\ldots\sum_{j_n=0}^{i_n}
\left\{ c_{i_1,\ldots,i_n,j_1,\ldots,j_n}^{({\rm I})}
\left(\prod_{m=1}^n\sqrt{I_m^{i_m}}\right) 
\cos\left[\sum_{m=1}^n(i_m-2j_m)\phi_m\right] \right .
\cr
& & \qquad\qquad\qquad\qquad\ \ +
\left. d_{i_1,\ldots,i_n,j_1,\ldots,j_n}^{({\rm I})}
\left(\prod_{m=1}^n\sqrt{I_m^{i_m}}\right)
\sin\left[\sum_{m=1}^n(i_m-2j_m)\phi_m\right] \right\}\ ,
\cr
}}
\label{eq:prototype_exp_act_ang}
\end{equation}
where $c_{i_1,\ldots,i_n,j_1,\ldots,j_n}^{({\rm I})}$ and
$d_{i_1,\ldots,i_n,j_1,\ldots,j_n}^{({\rm I})}$ are real
coefficients.

In principle, the form of the expansion~(\ref{eq:ham_BNF_deg_2}) of
the Hamiltonian $H^{({\rm I})}$ would be perfectly suitable to perform
the procedure constructing invariant tori near an elliptic equilibrium
point as it is described in~\cite{Loc-Gio-2000}. Unfortunately, in our
model of the planetary problem, the initial conditions
reported in Table~\ref{tab:freq_condiniz_sec_2D_SJSU} are far enough
from the equilibrium point to induce some effects of numerical
instability that are mainly due to the determination of the actions
translations (which will be properly defined in the following); this
can prevent the convergence of the algorithm constructing the KAM
torus. In order to circumvent such an obstruction, it is better to
adapt to the present context the normalization scheme introduced
in~\cite{Gab-Jor-Loc-2005}, where the previous and more usual approach
is refined so as to produce an algorithm that is effective also in a
region of the phase space not so close to the elliptic equilibrium
point. In order to make this work rather self-consistent, it is
convenient to first recall in subsect.~\ref{sbs:Kolm_norm_form} the
algorithm described in~\cite{Loc-Gio-2000}. Finally, in
subsect.~\ref{sbs:var_theme} we show the modifications that will
allow us to construct invariant tori in a wider neighbourhood of the
equilibrium point.

\subsection{Kolmogorov normalization near an elliptic
 equilibrium point}\label{sbs:Kolm_norm_form} The goal is to introduce
a suitable sequence of canonical transformations leading the
Hamiltonian~(\ref{eq:ham_BNF_deg_2}) in Kolmogorov normal
form\footnote{By a little abuse of notation we denote 
  by $\vet{\omega}$ the angular velocity vector, as usual in KAM
  theory. This should not be confused with the longitudes of perihelia, as usual in Celestial Mechanics.}, i.e.,
\begin{equation}
  H^{(\infty)}(\vet{p},\vet{q})=
  \vet{\omega}\cdot\vet{p}+\Oscr\big(\|\vet{p}\|^2\big)\ ,
\label{eq:H_infty}
\end{equation}
where $(\vet{p},\vet{q})\in\reali^n\times\toro^n$ are action-angle
coordinates; thus, the surface $\vet{p}=\vet{0}$ is invariant with
respect to the flow induced by $H^{(\infty)}$ and the motion over that
torus has angular velocities equal to the entries of a prescribed
vector $\vet{\omega}\in\reali^n$. The algorithm consists of a sequence of
canonical transformations that we describe in three separated steps.

\vskip 0.2truecm

\wideitem{(i)} {\it Birkhoff normalization up to a finite degree}

\nobreak\noindent We first determine a generating
function $B^{({\rm II})}$ by solving the equation
\begin{equation}
\sum_{j=1}^3\nu_j\,\frac{\partial B^{({\rm II})}}{\partial\phi_j}
+f_2^{({\rm I})}-\langle f_2^{({\rm I})}\rangle_{\vet{\phi}} = 0\ .
\label{eq:gen_Birkhoff_2}
\end{equation}
The expansion of the transformed Hamiltonian $H^{({\rm II})}= \exp
\lie{B^{({\rm II})}}\>H^{({\rm I})}$ can be written as
\begin{equation}
H^{({\rm II})}(\vet{I},\vet{\phi}) = \vet{\nu}\cdot\vet{I} +
f_{2}^{({\rm II})}(\vet{I})+\sum_{s\ge 2} f_{2s}^{({\rm II})}(\vet{I},\vet{\phi})\ ,
\label{eq:ham_Birkhoff_2}
\end{equation}
where the occurrence of the new normal form term $f_{2}^{({\rm
    II})}=\langle f_2^{({\rm I})}\rangle_{\vet{\phi}}$ has been
highlighted, by separating it from the series of the perturbing terms.
The recursive expression of $f_{2s}^{({\rm II})}$ as a function of
$B^{({\rm II})}$ and $f_{2l}^{({\rm I})}$ (with $l\le s$) can be
computed by just collecting the homogeneous polynomials having the
same degree in the square root of the actions. Thus, it is easy to
check that the functions $f_{2s}^{({\rm II})}$ are of the same type as
in~(\ref{eq:prototype_exp_act_ang}).

We stress that the Birkhoff normal form up to degree $2$ in the
actions is enough to start the following construction of the
Kolmogorov normal form. On the other hand, by performing the
Birkhoff normalization up to a degree higher than $3$, we can improve
the numerical stability of the calculation of the coefficients
appearing in the expansions generated by the algorithm. Here, we have
computed the Birkhoff normalization up to the third degree in
$\vet{I}$ which is good enough for our purposes, in the framework of
the model we are studying. Therefore, the final Hamiltonian is given
by $H^{({\rm III})} =\exp \lie{B^{({\rm III})}}\>H^{({\rm II})}$, that is
\begin{equation}
H^{({\rm III})}(\vet{I},\vet{\phi}) = \vet{\nu}\cdot\vet{I} +
f_{2}^{({\rm III})}(\vet{I})+
f_{4}^{({\rm III})}(\vet{I})+\sum_{s\ge 3} f_{2s}^{({\rm III})}(\vet{I},\vet{\phi})\ ,
\label{eq:ham_Birkhoff_3}
\end{equation}
where (a) the functions $f_{2s}^{({\rm III})}$ are homogeneous polynomials
of degree $2s+2$ in the square root of the actions $I$ and are of
type~(\ref{eq:prototype_exp_act_ang}); (b) the generating function
$B^{({\rm III})}$ is defined by the equation
\begin{equation}
\sum_{j=1}^3\nu_j\,\frac{\partial B^{({\rm III})}}{\partial\phi_j}
+f_4^{({\rm II})}-\langle f_4^{({\rm II})}\rangle_{\vet{\phi}} = 0\ ;
\label{eq:gen_Birkhoff_3}
\end{equation}
(c) $f_4^{({\rm III})}=\langle f_4^{({\rm II})}\rangle_{\vet{\phi}}\,$.
As a whole, the canonical transformation $\Bscr$ inducing the Birkhoff
normalization up to degree three in actions is explicitly given by
$\Bscr(\vet{I},\vet{\phi})=\exp \lie{B^{({\rm III})}}\,\circ\,\exp
\lie{B^{({\rm II})}}\>(\vet{I},\vet{\phi})\,$, indeed it is easy to
check that $H^{({\rm III})}(\vet{I},\vet{\phi})=H^{({\rm
    I})}\big(\Bscr(\vet{I},\vet{\phi})\big)$ using the exchange
theorem for Lie series.

\vskip 0.2truecm

\wideitem{(ii)} {\it Initial translation of the actions}

\nobreak\noindent The canonical transformation
$(\vet{I},\vet{\phi})=\Tgot_{\vet{I}^*}(\vet{p},\vet{q})$ performing
the initial translation of the actions is of type
\begin{equation}
I_j=p_j+I_j^*\ ,\qquad \varphi_j=q_j\ ,
\qquad j=1,\,\ldots,\,n\ .
\label{eq:trasf_trasl}
\end{equation}
Let us recall that we are constructing an invariant torus with a fixed
frequency vector $\vet{\omega}$.  Following~\cite{Loc-Gio-2000}, the
initial translation can be determined in such a way that, {\it in the
  integrable approximation}, the quasi-periodic motions on the
invariant torus $(\vet{p}=\vet{0}\,,\,\vet{q}\in\toro^n)$ have angular
frequencies $\vet{\omega}$. Therefore, we determine the vector
$\vet{I}^*$ with positive components (recall the definition of the
canonical transformation
$(\vet{x},\vet{y})=\Ascr(\vet{I},\vet{\phi})$) as the nearest to the
origin solution of the equations
\begin{equation}
  \nu_j+\frac{\partial f_2^{({\rm III})}}{\partial I_j}(\vet{I})+
  \frac{\partial f_4^{({\rm III})}}{\partial I_j}(\vet{I})=
  \omega_j\ , \qquad j=1,\,\ldots,\,n\ .
\label{eq:def_trasl_1}
\end{equation}
\noindent We can write the expansion of
$H^{({\rm IV})}(\vet{p},\vet{q})=
H^{({\rm III})}\big(\Tgot_{\vet{I}^*}(\vet{p},\vet{q})\big)$ as
\begin{equation}
H^{({\rm IV})}(\vet{p},\vet{q})= \vet{\omega}\cdot\vet{p}+\sum_{s\ge
      0}\sum_{l\ge 0}f_l^{({\rm IV},s)}(\vet{p},\vet{q}) \ ,
\label{eq:ham_trasl}
\end{equation}
where, for $l\ge 0$ and $s\ge 0\,$, $f_l^{({\rm IV},s)}$ is a
homogeneous polynomial of degree $l$ in the actions $\vet{p}$ and a
trigonometric polynomial of degree either $2s$ or $2s-1$ in the angles
$\vet{q}\,$. For short, let us introduce the symbol $\Pscr_{l,2s}\,$,
which denotes the set of functions that are homogeneous polynomials of
degree $l$ in the actions and trigonometric polynomials of degree at
most $2s$ in the angles, thus, $f_l^{({\rm IV},s)}\in\Pscr_{l,2s}\,$.
Moreover, using the Cauchy inequalities, one easily sees that the size
(of any suitable norm) of $f_l^{({\rm IV},s)}$ can be estimated with
an upper bound that is essentially proportional to the $s$-th power
of the ratio of $\|\vet{I}^*\|$ over the analytic radius of
convergence of $H^{({\rm III})}$, and it is inversely proportional to
the $l$-th power of the minimum component of vector
$\vet{I}^*$. Therefore, $\vet{I}^*$ plays a major role in the
convergence of the expansions, because it is proportional to what is
commonly identified as the small parameter of the KAM theory and it rules the
radius of convergence for the actions $\vet{p}$.  At this point, we
want emphasize that we have some freedom in the crucial choice of the
initial translation vector $\vet{I}^*$, as it will be discussed in
sect.~\ref{sbs:var_theme}.

\vskip 0.2truecm

\wideitem{(iii)} {\it The standard Kolmogorov normalization algorithm}

\nobreak\noindent
Let us describe the generic $r$-th step of the Kolmogorov
normalization algorithm. We begin with a Hamiltonian of the type
\begin{equation}
H^{(r-1)}(\vet{p},\vet{q})=
\vet{\omega}\cdot\vet{p}+\sum_{s\ge 0}\sum_{l\ge 0}f_l^{(r-1,s)}(\vet{p},\vet{q})
\ ,
\label{eq:H^r-1}
\end{equation}
where $f_l^{(r-1,s)}\in\Pscr_{l,2s}\,$, for $l\ge 0$ and $s\ge
0\,$.  To fix ideas, we can start with $r=2$ defining
$H^{({1})}=H^{({\rm IV})}$.  Since we point to a Hamiltonian of
type~(\ref{eq:H_infty}), we must remove the main perturbing terms of
degree $0$ and $1$ in the actions.  We will proceed in two separate
steps.  We first remove part of the unwanted terms via a canonical
transformation with generating function
$\chi_1^{(r)}(\vet{q})=X^{(r)}(\vet{q})+\vet{\xi}^{(r)}\cdot\vet{q}$
(being $\vet{\xi}^{(r)}\in\reali^n$). Thus, we solve with respect to
$X^{(r)}(\vet{q})$ and $\vet{\xi}^{(r)}\,$ the equations
\begin{equation}
\sum_{j=1}^n\omega_j\frac{\partial\,X^{(r)}}{\partial q_j}(\vet{q})+
\sum_{s=1}^r f^{(r-1,s)}_0(\vet{q})=0
\ ,
\quad
C^{(r)}\vet{\xi}^{(r)}\cdot\vet{p}+f^{(r-1,0)}_1(\vet{p})=0\ ,
\label{eq:chi_1}
\end{equation}
where the $n\times n$ matrix $C^{(r)}$ is defined by the equation
$\frac{1}{2}C^{(r)}\vet{p}\cdot\vet{p}=f_2^{(r-1,0)}(\vet{p})\,$. A
unique solution satisfying $\langle X^{(r)}\rangle_{\vet{q}}=0$ exists
if the frequencies $\vet{\omega}$ are non-resonant up to order $2r$,
$\vet{k}\cdot\vet{\omega}\neq 0\,$, with $\vet{k}\in\interi^n$ such
that $0<|\vet{k}|\le 2r$, and if $\det C^{(r)}\neq 0\,$. We must now
give the expressions of the functions $\hat f_l^{(r,s)}$ appearing in
the expansion of the new Hamiltonian
\begin{equation}
\hat H^{(r)}(\vet{p},\vet{q})=
\vet{\omega}\cdot\vet{p}+\sum_{s\ge 0}\sum_{l\ge 0}\hat f_l^{(r,s)}(\vet{p},\vet{q})
\ ,
\label{eq:hatH^r}
\end{equation}
where $\hat H^{(r)}=\exp\lie{\chi_1^{(r)}}H^{(r-1)}$. To this aim, we
will redefine many times the same quantity without changing the
symbol. This is made by mimicking the {\tt C} programming language, by using the
notation $a\pluseq b$ which means that the previously defined quantity
$a$ is redefined as $a=a+b\,$. Therefore, we initially define
\begin{equation}
\hat f_l^{(r,s)}=f_l^{(r-1,s)}(\vet{p},\vet{q})\ ,
\quad\ {\rm for}\ l\ge 0\ {\rm and}\ s\ge 0\ .
\label{eq:hatf_l^rs_def_1}
\end{equation}
To take into account the Poisson bracket of the generating function
with $\vet{\omega}\cdot\vet{p}$, we put
\begin{equation}
\hat f_0^{(r,0)}\pluseq \vet{\omega}\cdot\vet{\xi}^{(r)}\ ,
\qquad
\hat f_0^{(r,s)}=0
\quad\ {\rm for}\ 1\le s\le r\ .
\label{eq:hatf_l^rs_def_2}
\end{equation}
Then, we consider the contribution of the terms generated by the Lie
series applied to each function $f_l^{(r-1,s)}$ as
\begin{equation}
\hat f_{l-j}^{(r,s+jr)}\pluseq
\frac{1}{j!}\lie{\chi_1^{(r)}}^jf_l^{(r-1,s)}
\quad\ {\rm for}\ l\ge 1\,,\ s\ge 0\ {\rm and}\ 1\le j\le l\ .
\label{eq:hatf_l^rs_def_3}
\end{equation}
Looking at
formul\ae~(\ref{eq:hatf_l^rs_def_1})--(\ref{eq:hatf_l^rs_def_3}), one
can easily check that $\hat f_l^{(r,s)}\in\Pscr_{l,2s}$, for
$l\ge 0$ and $s\ge 0\,$.  We perform now a {\it ``reordering of
  terms''}, by moving the monomials appearing in the expansion of a
function $\hat f_l^{(r,s)}$ to another, in such a way that the
so redefined functions $\hat f_l^{(r,s)}$ are homogeneous polynomials of
degree $l$ in the actions and trigonometric polynomials of degree $2s$
or $2s-1$ in the angles, for $l\ge 0$ and $s\ge 0\,$.

In the second part of the $r$-th step of the Kolmogorov's
normalization algorithm, by using another canonical transformation, we
remove the part of the perturbation up to the order of magnitude $r$
that actually depends on the angles and it is linear in the actions.
Thus, we solve with respect to $\chi_2^{(r)}(\vet{p},\vet{q})$ the equation
\begin{equation}
\sum_{j=1}^n\omega_j\frac{\partial\,\chi_2^{(r)}}{\partial
  q_j}(\vet{p},\vet{q})+ \sum_{s=1}^r \hat f^{(r,s)}_1(\vet{p},\vet{q})=0 \ ,
\label{eq:chi_2}
\end{equation}
where again the solution exists and it is unique if $\langle
\chi_2^{(r)}\rangle_{\vet{q}}=0$ and the frequencies $\vet{\omega}$
are non-resonant up to order $2r$. Analogously to what we have done
above, we now provide the expressions of the functions $f_l^{(r,s)}$
appearing in the expansion of the new Hamiltonian
\begin{equation}
H^{(r)}(\vet{p},\vet{q})= \vet{\omega}\cdot\vet{p}+\sum_{s\ge
  0}\sum_{l\ge 0} f_l^{(r,s)}(\vet{p},\vet{q}) \ ,
\label{eq:H^r}
\end{equation}
where $H^{(r)}=\exp\lie{\chi_2^{(r)}}\hat H^{(r)}$. We initially define
\begin{equation}
f_l^{(r,s)}=\hat f_l^{(r,s)}(\vet{p},\vet{q})
\quad\ {\rm for}\ l\ge 0\ {\rm and}\ s\ge 0\ .
\label{eq:f_l^rs_def_1}
\end{equation}
In order to take into account the contribution of the terms generated
by the Lie series applied to $\vet{\omega}\cdot\vet{p}\,$, we put
\begin{equation}
f_1^{(r,jr)}\pluseq
-\frac{1}{j!}\lie{\chi_2^{(r)}}^{j-1}
\left(\sum_{s=1}^r \hat f^{(r,s)}_1(\vet{p},\vet{q})\right)
\quad\ {\rm for}\ j\ge 1\ .
\label{eq:f_l^rs_def_2}
\end{equation}
Then, the contribution of the Lie series applied to the rest of the
Hamiltonian $\hat H^{(r)}$ implies that
\begin{equation}
f_{l}^{(r,s+jr)}\pluseq
\frac{1}{j!}\lie{\chi_2^{(r)}}^j\hat f_l^{(r,s)}
\quad\ {\rm for}\ l\ge 0\,,\ s\ge 0\ {\rm and}\ j\ge 1\ .
\label{eq:f_l^rs_def_3}
\end{equation}
Finally, we perform a new {\it ``reordering of terms''}, so that
at the end the functions $f_l^{(r,s)}\in\Pscr_{l,2s}$ appearing in the
expansion~(\ref{eq:H^r}) of the new Hamiltonian $H^{(r)}$ are 
again homogeneous polynomials of degree $l$ in the actions and 
trigonometric polynomials of degree $2s$ or $2s-1$ in the angles, 
for $l\ge 0$ and $s\ge 0\,$.

\noindent Let us recall that the canonical transformation $\Kscr^{(r)}$
inducing the Kolmogorov normalization up to the step $r$ is
explicitly given by
\begin{equation}
\Kscr^{(r)}(\vet{p},\vet{q})=\exp\lie{\chi_2^{(r)}}\,\circ\,\exp\lie{\chi_1^{(r)}}
\,\circ\,\ldots\,\exp\lie{\chi_2^{(2)}}\,\circ\,\exp\lie{\chi_1^{(2)}}
\>(\vet{p},\vet{q})\ .
\label{eq:trasf_Kolmogorov}
\end{equation}
This conclude the $r$-th step of the algorithm that can be 
further iterated.

\subsection{The modified algorithm constructing the Kolmogorov 
normal form}\label{sbs:var_theme}

As the prediction of the translation vectors $\vet{\xi}^{(r)}$ given
by (\ref{eq:chi_1}) can be affected by large errors, we have split the
standard Kolmogorov normalization algorithm in two separate steps.
First, we iterate for a fixed number of steps the normalization
algorithm by setting to zero the translation vectors
$\vet{\xi}^{(r)}$.  This procedure is reminiscent of Arnold's proof
of the KAM theorem (\cite{Arnold-1963} and see
also~\cite{San-Loc-Gio-2011} and~\cite{Gio-Loc-San-2014}, where such a
modification of Kolmogorov normalization algorithm has been recently
adapted so as to approximate elliptic lower dimensional tori and to prove
their existence).  Therefore, under mild theoretical assumptions, such
a partial normalization procedure can still converge  to a Hamiltonian in Kolmogorov normal form related to a
vector of angular frequencies $\vet{\omega}^*$ different from
$\vet{\omega}$.  In practice, we perform just a finite number of steps
of this partial normalization. Then, using this intermediate
Hamiltonian $H\simeq\vet{\omega}^*\cdot\vet{p}
+\Oscr\big(\|\vet{p}\|^2\big)$ as the initial one, we restart a
complete standard Kolmogorov normalization algorithm now including
the translation vectors $\vet{\xi}^{(r)}$ defined
in~(\ref{eq:chi_1}).  This splitting of the normalization algorithm in
two separate steps becomes advantageous if, after the first step, the
frequencies $\vet{\omega}^*$ are sufficiently close to $\vet{\omega}$
and, thus, the translation vectors $\vet{\xi}^{(r)}$ are small
enough.  Let us now remark that the frequencies $\vet{\omega}^*$,
related to the intermediate Hamiltonian
$H\simeq\vet{\omega}^*\cdot\vet{p} +\Oscr(\|\vet{p}\|^2)$, depend on
the initial translation vector $\vet{I}^*$ of the canonical
transformation (\ref{eq:trasf_trasl}).  Therefore, we try to choose
$\vet{I}^*$ in such a way that the frequency vector $\vet{\omega}^*$
is as close as possible to $\vet{\omega}$.

We now provide a detailed description of our algorithm, that has been
formulated so as to reproduce the main ideas explained in the previous
heuristic discussion.  Let us start from a system of
the type described by the Hamiltonian $H^{({\rm I})}$
in~(\ref{eq:ham_BNF_deg_2}). Then, let us carry out the following
steps.
\begin{enumerate}[(a)]
  \setlength{\itemsep}{0pt}
  \setlength{\parskip}{0pt}
  \setlength{\parsep}{0pt}
  \setlength{\topsep}{0pt}
    
\item Perform the canonical transformation $\Bscr$ that
realizes the Birkhoff normalization up to a finite degree, as
described at point (i) of sect.~\ref{sbs:Kolm_norm_form}.

\item Determine a good initial translation vector
$\vet{{\hat I}}$ by proceeding as follows.
  \begin{enumerate}[(b1)]
    \setlength{\partopsep}{0pt}
  \setlength{\itemsep}{0pt}
  \setlength{\parskip}{0pt}
  \setlength{\parsep}{0pt}
  \setlength{\topsep}{0pt}
\item Let us refer to the initial conditions as
$(\vet{I}_0,\vet{\phi}_0)$; calculate their values in the new
coordinates, say
$(\vet{I}_0^*,\vet{\varphi}_0^*)=\Bscr(\vet{I}_0,\vet{\varphi}_0)$. Then,
perform the initial translation of the actions, as at point (ii) of
sect.~\ref{sbs:Kolm_norm_form}, by replacing $\vet{I}^*$ with
$\vet{I}_0^*\,$.

\item Let us perform the Kolmogorov
normalization algorithm up to a fixed $R^{\prime}$-th step, as at
point (iii) of sect.~\ref{sbs:Kolm_norm_form}, starting from
$H^{({1})}=H^{({\rm IV})}$, but putting the translation vectors
$\vet{\xi}^{(r)}=\vet{0}\,$, for $r=2,\,\ldots,\,R^{\prime}\,$.
Let us define $\vet{\omega}_0^*$ in such a way that
$\vet{\omega}_0^*\cdot\vet{p}=\vet{\omega}\cdot\vet{p}+f_1^{(0,r+1)}(\vet{p})$,
with $f_1^{(0,r+1)}$ as obtained at the end of such
procedure. $R^{\prime}$ is a fixed integer parameter that is selected
sufficiently large to allow the convergence of the whole algorithm,
but also taking into consideration the computational resources
available.

\item Let us improve our choice of
$\vet{I}_0^*$, by approximating numerically (by the finite differences
method) the Jacobian matrix $\Jscr_{\vet{I}_0^*}$ of the function
$\vet{\omega}^*(\vet{I}_0^*)$ and then solving the linear equation
$\Jscr_{I_0^*}(\vet{{\hat
    I}}-\vet{I}^*)=\vet{\omega}-\vet{\omega}_0^*$ in the unknown
$\vet{{\hat I}}$.
\end{enumerate}
\item Perform the translation of the actions as at point (ii),
sect.~\ref{sbs:Kolm_norm_form}, replacing $\vet{I}^*$ with
$\vet{{\hat I}}$.

\item Let us perform again the Kolmogorov normalization
algorithm without translations, as at step (${\rm b}_{2}$). In what
follows we will denote as $\{\Hgot^{(r)}\}_{r=1}^{R^{\prime}}\,$,
$\{\Xgot_1^{(r)},\Xgot_2^{(r)}\}_{r=2}^{R^{\prime}}$ and
$\{\Kgot^{(r)}\}_{r=2}^{R^{\prime}}$ the obtained finite sequences of
the Hamiltonians, the generating functions and the canonical
transformations, respectively; so that $\Hgot^{({1})}=H^{({\rm IV})}$,
$\Hgot^{(r)}=\exp\lie{\Xgot_2^{(r)}}\big(\exp\lie{\Xgot_1^{(r)}}\Hgot^{(r-1)}\big)$
and $\Hgot^{({r})}=\Hgot^{({1})}\circ\Kgot^{(r)}$, for
$r=2,\,\ldots,\,R^{\prime}\,$. Therefore, $\Kgot^{(r)}$ is explicitly
given by a formula analogous to~(\ref{eq:trasf_Kolmogorov}), by
replacing the symbols $\Kscr$ and $\chi$ with $\Kgot$ and $\Xgot$,
respectively.

\item Let us perform the standard Kolmogorov normalization
algorithm, as at point (iii) of sect.~\ref{sbs:Kolm_norm_form} (with
the translation vectors $\vet{\xi}^{(r)}$ given by~(\ref{eq:chi_1})),
starting from $H^{({1})}=\Hgot^{(R^{\prime})}$.

\end{enumerate}

\vskip 0.2truecm

Let us remark that steps (b1)--(b3) of the previous procedure
can be iterated more than once, in order to refine the calculation of
the initial translation vector $\vet{{\hat I}}\,$; however, from a
practical point of view, a single execution of the steps (b1)--(b3)
is often enough to successfully perform the subsequent Kolmogorov
normalization algorithms, in the sense that one can clearly appreciate
their numerical convergence. In particular, this applies also in the
case of our model of the planetary problem we are studying.

The knowledge of the normal form (and of the normalizing
transformations) allows to explicitly implement a semi-analytic
procedure producing the integration of the equations of motion. For
instance, if we consider the coordinates $(\vet{x},\vet{y})$
introduced at the very beginning of the present section (where the
canonical transformation $\Dscr$ was defined) and the action-angle
variables $(\vet{p},\vet{q})$ of the normal form, we have that
\begin{equation}
\label{eq:semianalytical-integration}
\vcenter{\openup1\jot\halign{
 \hbox to 25 ex{\hfil $\displaystyle {#}$\hfil}
 &\hbox to 12 ex{\hfil $\displaystyle {#}$\hfil}
 &\hbox to 32 ex{\hfil $\displaystyle {#}$\hfil}
 &\hbox to 3 ex{\hfil $\displaystyle {#}$\hfil}
 \cr
 \big( \vet{x}(0) \,,\, \vet{y}(0) \big)
 &\build{\longrightarrow}_{}^{
 {{\displaystyle \left(\Cscr^{(\infty)}\right)^{-1}} \atop \phantom{0}}}
 &\left({{\displaystyle \vet{p}(0)=\vet{0}} \,,\, {\displaystyle \vet{q}(0)}}\right)
 \cr
 & &\Big\downarrow &,
 \cr
 \big( \vet{x}(t) \,,\, \vet{y}(t) \big)
 &\build{\longleftarrow}_{}^{
 {{\displaystyle \Cscr^{(\infty)}} \atop \phantom{0}}}
 &\left({{\displaystyle \vet{p}(t)=\vet{p}(0)}
 \,,\, {\displaystyle \vet{q}(t)= \vet{q}(0)+\vet{\omega} t}}\right)
 \cr
 }}
\end{equation}
with $\Cscr^{(\infty)}=\lim_{r\to\infty}\Cscr^{(r)}$ and
$\Cscr^{(r)}=\Ascr\circ\Bscr\circ\Tgot_{\vet{{\hat I}}}\circ
\Kgot^{(R^{\prime})}\circ\Kscr^{(r)}$, where the canonical
transformation $\Ascr$ is defined at the very beginning
of the present section, while $\Bscr\,$, $\Tgot_{\vet{{\hat I}}}\,$,
$\Kgot^{(R^{\prime})}$ and $\Kscr^{(r)}$ are determined at points~(a),
(c), (d) and~(e) of the algorithm described above, respectively.

\subsection{Application to the secular model}\label{sbs:appl_KAM_2_sjsu_2D_sec}
We have checked our implementation by comparing the numerical
integration of the flow induced by the Hamiltonian $H^{(\Dscr)}$
(written in~\eqref{Ham^0}) against the results from the normal form,
by means of the scheme~(\ref{eq:semianalytical-integration}).  The
input data needed to start the execution of the algorithm constructing
the Kolmogorov normal form are the coefficients appearing in the
expansion of $H^{(\Dscr)}$ and the frequency vector $\vet{\omega}$
identifying the KAM torus we are looking for. In order to determine
$\vet{\omega}$, we preliminary perform a long-time numerical
integration of the motion law $t\mapsto\big( \vet{x}(t) \,,\,
\vet{y}(t)\big)$ starting from the initial conditions reported in
Table~\ref{tab:freq_condiniz_sec_2D_SJSU}. Then, we apply the
frequency analysis method (see~\cite{Laskar-03} and references
therein) to the signals $x_j(t)+{\rm i}y_j(t)\,$, with
$j=1,\,2,\,3\,$. The corresponding fundamental frequencies detected by
that numerical method provide us the wanted values

\begin{equation}
\vcenter{\openup1\jot\halign{
 \hbox {\hfil $\displaystyle {#}$}
&\hbox {\hfil $\displaystyle {#}$\hfil}
&\hbox {$\displaystyle {#}$\hfil}\cr
  \omega_{1} &= &-1.51665408389554804\,\times 10^{-4}\ ,
  \qquad
  \omega_{2} = -2.05981220762083458\,\times 10^{-5}\ ,
  \cr
  & &\qquad\qquad\qquad\qquad
  \omega_{3} = -1.16008414414439544\,\times 10^{-5}\ .
  \cr
}}
\end{equation}

\begin{figure}[t]
\centerline{\includegraphics[width=155mm]{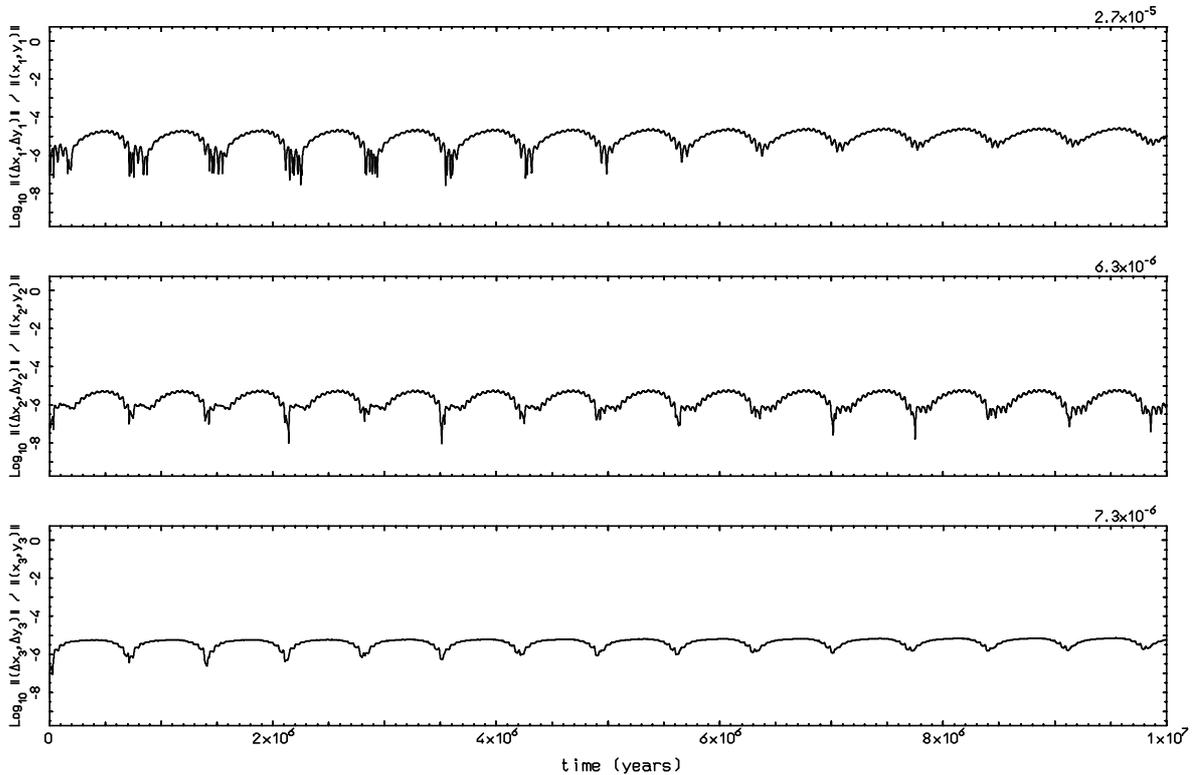}}
\caption{Test on the reliability of the construction of the
  Kolmogorov normal form for the planar secular model of the system
  including the Sun, Jupiter, Saturn and Uranus. All the boxes report
  (in a semi-log scale) the time dependency of the discrepancies about
  the computed values of the canonical coordinates $(\vet{x},\vet{y})$
  between a numerical integration and the semi-analytic one, based on
  an approximation of the
  scheme~(\ref{eq:semianalytical-integration}). The maximal value of
  the ordinate of the plotted points is reported near to the
  top-right corner of each box.}
  \label{fig:dis_conf_kam}
\end{figure}

\noindent
Of course, formula~\eqref{eq:semianalytical-integration} refers to an
ideal scheme, an infinite sequence of canonical transformations cannot
be performed on a computer; thus, we limited ourselves to approximate
the exact solution by replacing $\Cscr^{(\infty)}$ with $\Cscr^{(25)}$
and putting also $R^{\prime}=25$. Both the (approximate) normalizing
transformations $\Ascr\circ\Bscr\circ\Tgot_{\vet{{\hat I}}}\circ
\Kgot^{(25)}\circ\Kscr^{(25)}$ and its inverse have been computed by
using our software for algebraic manipulations. The results of a
long-time comparison (lasting for $10^7$ years) are summarized in
Figure~\ref{fig:dis_conf_kam}: the relative error on the determination
of the two-dimensional vectors $\big(x_j(t)\,,\,y_j(t)\big)$ is always smaller than
$0.003\,$\%, for $j=1,\,2,\,3$. We stress that the accuracy of the
semi-analytic solution is mainly affected by the unavoidable
truncation rules that are applied to the expansions. Let us shortly
describe them all together as follows: we have put the same limits on
the partial Birkhoff normalization as those holding for $\Hscr^{({\rm
    sec})}$, i.e., the Hamiltonians $H^{({\rm
    I})},\,\ldots\,\,H^{({\rm IV})}$ have been truncated up to degree
18 in the square root of the actions $\vet{I}\,$; during both the
Kolmogorov normalization algorithms the expansions of the
Hamiltonians were limited to the fifth degree in actions $\vet{p}$ and
up to a maximal trigonometric degree in $\vet{q}$ equal to 50. The
truncation rules applied to the normalizing transformations are
analogous to those for the Hamiltonians.

Taking into account all the factors limiting the accuracy of our
approximation of the semi-analytic integration
scheme~\eqref{eq:semianalytical-integration}, we consider that the
agreement with the numerical integration is excellent. In our opinion,
this is a strong validation of all the procedure constructing the
Kolmogorov normal form. We also think that, with a relevant
additional effort, our approach could be translated in a
computer-assisted proof of existence of such an invariant torus, in a
similar way to what was done in~\cite{Loc-Gio-2000}
and~\cite{Gab-Jor-Loc-2005}. However, this goes far beyond the scopes
of the present work.

\section{Effective stability in the neighborhood of the KAM torus}
\label{sec:nekh}

We follow the approach described in~\cite{Mor-Gio-1995} and already
applied to another planetary problem in~\cite{Gio-Loc-San-2009}. We
focus on the Kolmogorov normal form~\eqref{eq:H_infty}, which is
related to the locally unique invariant surface, that is travelled by
quasi-periodic motion characterized by the angular velocity vector
$\vet{\omega}$. $H^{(\infty)}$ is now reconsidered as a new starting
point for performing a Birkhoff normalization. Such a further
constructive procedure finally aims to provide bounds on the diffusion
from that torus having angular velocity equal to $\vet{\omega}$; this
is made by giving a lower estimate on the escape time $T(\rho)$. Here,
$\rho$ is the distance (as measured in actions) from the initial torus
related to $\vet{\omega}$ and plays the role of the small parameter in
the construction of the Birkhoff normal form.

\subsection{Formal algorithm constructing the Birkhoff normal form close to the KAM torus}\label{sbs:alg-local-Birk}
In order to properly describe the new normalization algorithm, it is
convenient to introduce the new symbol $\Hscr^{(2,0)}=H^{(\infty)}$ for the
(re)starting Hamiltonian and we explicitly write its expansion as 
\begin{equation}
  \Hscr^{(2,0)}(\vet{p},\vet{q})= \vet{\omega}\cdot\vet{p}+
  \sum_{l\ge 2}\sum_{s\ge 0} f_l^{(2,0;s)}(\vet{p},\vet{q}) \ ,
\label{eq:Hscr^2,0}
\end{equation}
being $f_l^{(2,0;s)}\in\Pscr_{l,2s}\,$. Unfortunately, in order to
obtain final estimates that are good enough, we cannot simply
prescribe that, as a rule to perform the $l$-step, all the
perturbative terms having the same $l$ degree in actions must be
removed at the same time; this would somehow natural but too
rough. Therefore, we need to design a normalization procedure based on
two indexes. Let us describe the generic $(r,\rgot)$ step, which aims
to introduce the new Hamiltonian
$\Hscr^{(r,\rgot)}=\exp\lie{\Bgot^{(r,\rgot)}}\>\Hscr^{(r,\rgot-1)}$,
where the expansion of the previous Hamiltonian reads
\begin{equation}
  \Hscr^{(r,\rgot-1)}(\vet{p},\vet{q})= \vet{\omega}\cdot\vet{p}
  +\sum_{l=2}^{r} f_l^{(r,\rgot-1;0)}(\vet{p})
  +\sum_{s>\rgot-1} f_r^{(r,\rgot-1;s)}(\vet{p},\vet{q})
  +\sum_{l>r}\sum_{s\ge 0} f_l^{(r,\rgot-1;s)}(\vet{p},\vet{q}) \ ,
\label{eq:Hscr^r,rgot-1}
\end{equation}
being, again, $f_l^{(r,\rgot-1;s)}\in\Pscr_{l,2s}\,$.  The generating
function is determined so as to remove the main perturbing term among
those having $r$ degree with respect to the actions $\vet{p}$; in view
of the Fourier decay, this is $f_r^{(r,\rgot-1;r)}$, thus we set the
homological equation
\begin{equation}
\sum_{j=1}^n\omega_j\,\frac{\partial \Bgot^{(r,\rgot)}}{\partial q_j}
+f_r^{(r,\rgot-1;\rgot)}-\langle f_r^{(r,\rgot-1;\rgot)}\rangle_{\vet{q}} = 0\ ,
\label{eq:gen_Bgot^r,rgot}
\end{equation}
that must be obviously solved with respect to
$\Bgot^{(r,\rgot)}\in\Pscr_{r,2\rgot}\,$. We proceed in an analogous
way to what has been done in subsect.~\ref{sbs:Kolm_norm_form}, in
order to define all the terms appearing in the new expansion
\begin{equation}
  \Hscr^{(r,\rgot)}(\vet{p},\vet{q})= \vet{\omega}\cdot\vet{p}
  +\sum_{l=2}^{r} f_l^{(r,\rgot;0)}(\vet{p})
  +\sum_{s>\rgot} f_r^{(r,\rgot;s)}(\vet{p},\vet{q})
  +\sum_{l>r}\sum_{s\ge 0} f_l^{(r,\rgot;s)}(\vet{p},\vet{q}) \ .
\label{eq:Hscr^r,rgot}
\end{equation}
First, we initially set
\begin{equation}
f_l^{(r,\rgot;s)}=f_l^{(r,\rgot-1;s)}(\vet{p},\vet{q})
\quad\ {\rm for}\ l\ge 2\ {\rm and}\ s\ge 0\ .
\label{eq:f_l^rrgots_def_1}
\end{equation}
Accordingly to the homological
equation~\eqref{eq:gen_Bgot^r,rgot}, we redefine
the normal forms term of $r$ degree in $\vet{p}$ as
\begin{equation}
  f_r^{(r,\rgot;0)}\pluseq \langle f_r^{(r,\rgot-1;\rgot)}\rangle_{\vet{q}}\ ,
  \qquad
  f_r^{(r,\rgot;\rgot)}=0\ .
\label{eq:f_l^rrgots_def_2}
\end{equation}
In order to take into account the contribution of all the terms (but
the first two) that are generated by the Lie series applied to
$\vet{\omega}\cdot\vet{p}$, we put
\begin{equation}
f_{j(r-1)+1}^{(r,\rgot;j\rgot)}\pluseq
-\frac{1}{j!}\lie{\Bgot^{(r,\rgot)}}^{j-1}
\left(f_r^{(r,\rgot-1;\rgot)}-\langle f_r^{(r,\rgot-1;\rgot)}\rangle_{\vet{q}}\right)
\quad\ {\rm for}\ j\ge 2\ .
\label{eq:f_l^rrgots_def_3}
\end{equation}
Then, the contribution of the Lie series applied to the rest of the
Hamiltonian $\Hscr^{(r,\rgot-1)}$ implies that
\begin{equation}
f_{l+j(r-1)}^{(r,\rgot;s+j\rgot)}\pluseq
\frac{1}{j!}\lie{\Bgot^{(r,\rgot)}}^j f_l^{(r,\rgot-1;s)}
\quad\ {\rm for}\ l\ge 2\,,\ s\ge 0\ {\rm and}\ j\ge 1\ ,
\label{eq:f_l^rrgots_def_4}
\end{equation}
where some of the new summands actually have not any influence,
because $f_l^{(r,\rgot-1;s)}=0$ if $2\le l\le r$ and $s>0$ or $l=r$
and $1\le s<\rgot$ (in agreement with
equation~\eqref{eq:Hscr^r,rgot-1}).  All the
prescriptions~\eqref{eq:f_l^rrgots_def_1}--\eqref{eq:f_l^rrgots_def_4},
have been settled so as to preserve the rules about the classes of
functions, so that $f_l^{(r,\rgot;s)}\in\Pscr_{l,2s}\,$, as
it can be easily checked by induction. Since the structure of the
expansion~\eqref{eq:Hscr^r,rgot} is coherent with that
in~\eqref{eq:Hscr^r,rgot-1}, the procedure can be iterated at the next
$(r,\rgot+1)$ step.

In order to start the elimination of the perturbing terms of higher
degree in~$\vet{p}\,$, we simply put
$\Hscr^{(r+1,0)}=\lim_{\rgot\to\infty}\Hscr^{(r,\rgot)}$. Since the
expansion~\eqref{eq:Hscr^2,0} of the initial Hamiltonian
$\Hscr^{(2,0)}$ is identical to that in~\eqref{eq:Hscr^r,rgot-1} when
$r=2$ and $\rgot=1$, then the sequence of the generating functions in
our normalization scheme can be ideally represented as follows:
$\Bgot^{(2,1)},\> \Bgot^{(2,2)},\> \ldots\>
\Bgot^{(2,\infty)},\Bgot^{(3,1)},\>\ldots\>
\Bgot^{(3,\infty)},\Bgot^{(4,1)},\>\ldots\>$

\subsection{Iterative scheme of estimates}\label{sbs:iter-estimates}
Although explicit computations according to the previous formal
algorithm are feasible (see~\cite{Gio-Loc-San-2009}), the expansions
of the functions introduced by that procedure become quickly so
cumbersome, that it is hard to deal with them for any algebraic
manipulator when the truncation orders are increased with respect to
both the actions and the angles. Here, we limit ourselves to produce a
rather rough lower bound on the diffusion time, in order to
drastically reduce the computational difficulty. This is mainly made
by iterating a scheme of estimates involving just the norm for each of
those functions, instead of computing all their expansions. In
practice, for any generic function $g\in\Pscr_{l,2s}\,$, we define its
norm as
\begin{equation}
  \|g\|=\sum_{i_1+\ldots+i_n=l}\,\sum_{|\vet{k}|\le 2s}
  \left| c_{i_1,\ldots,i_n,k_1,\ldots,k_n} \right|\ ,
  \label{eq:def-norma}
\end{equation}
being $\{c_{i_1,\ldots,i_n,k_1,\ldots,k_n}\}_{\vet{i},\vet{k}}$ the
{\it finite} set of coefficients appearing in the corresponding
Taylor-Fourier series
$$
g(\vet{p},\vet{q})=\sum_{i_1+\ldots+i_n=l}\,\sum_{|\vet{k}|\le 2s}
c_{i_1,\ldots,i_n,k_1,\ldots,k_n}\,p_1^{i_1}\ldots p_n^{i_n}
\,{\scriptstyle{{\displaystyle{\sin}}\atop{\displaystyle{\cos}}}}
\big(\vet{k}\cdot\vet{q}\big)\ ,  
$$
where the notation
${\scriptstyle{{\displaystyle{\sin}}\atop{\displaystyle{\cos}}}}$
means that either the sine or the cosine can appear and, for any fixed
harmonic $\vet{k}$, the choice is made unique according to the
following usual criterion: if there is an index $1\le j\le n$ such
that $k_j<0$ and $k_1=\ldots=k_{j-1}=0$ then the sine function
appears, otherwise the cosine.

Our iterative scheme of estimates mainly aims to provide
a set of computational rules to determine a sequence of majorants
$\Fscr_l^{(r,\rgot;s)}$ such that
\begin{equation}
  \|f_l^{(r,\rgot;s)}\|\le \Fscr_l^{(r,\rgot;s)}
  \qquad{\rm for}\ l\ge 2,\ s\ge 0\ .
  \label{diseq:stima-generale-passo-rrgot}
\end{equation}
Let us recall that the induction must be started from
$\Hscr^{(2,0)}=H^{(\infty)}$. To fix the ideas, let us refer to the
explicit calculation of the Kolmogorov normal form and its
truncation rules, as they have been described in
subsect.~\ref{sbs:appl_KAM_2_sjsu_2D_sec}. Therefore,
$\Fscr_l^{(2,0;s)}$ can be computed for $2\le l\le 5$ and $0\le s
\le 25$, by simply using the definition~\eqref{eq:def-norma} of the
norm, while for all the remaining initial majorants we put
$\Fscr_l^{(2,0;s)}=0$ when $l>5$ or $s\ge 25$.

We now describe how the iteration of the estimates works in the case
of the generic $(r,\rgot)$ constructive step of the Birkhoff normal
form. By induction hypothesis, let us suppose to know the upper bounds
related to the previous step, that are $\|f_l^{(r,\rgot-1;s)}\|\le
\Fscr_l^{(r,\rgot-1;s)}$ for $l\ge 2,\ s\ge 0$. First, we estimate
the generating function so that
\begin{equation}
  \|\Bgot^{(r,\rgot)}\|\le \Gscr^{(r,\rgot)}\ ,
  \qquad {\rm with}\quad
  \Gscr^{(r,\rgot)}=\frac{\Fscr_r^{(r,\rgot-1;\rgot)}}{\alpha_\rgot}\ ,
  \label{diseq:iter-estimate-1}
\end{equation}
being\footnote{From a practical point of view, for not too large
  values of $\rgot$, a lower bound on $\min_{0<|\vet{k}|\le
    2\rgot}|\vet{k}\cdot\vet{\omega}|$ can be rigorously computed by
  using interval arithmetics, while the asymptotic behavior can be
  estimated by using the Diophantine inequality
  $|\vet{k}\cdot\vet{\omega}|\ge\gamma/|\vet{k}|^{\tau}$, for suitable
  $\gamma>0$ and $\tau\ge n-1$. For instance, \cite{Cel-Fal-Loc-2004}
  provides a general prescription rule to determine a vector
  satisfying the Diophantine property with the optimal value for
  $\tau=n-1$.}  $\alpha_\rgot=\min_{0<|\vet{k}|\le
  2\rgot}|\vet{k}\cdot\vet{\omega}|$. In order to take into account
the
prescriptions~\eqref{eq:f_l^rrgots_def_1}--\eqref{eq:f_l^rrgots_def_2},
we initially put
\begin{equation}
  \Fscr_l^{(r,\rgot;s)}=\Fscr_l^{(r,\rgot-1;s)} \ ,
  \qquad{\rm for}\ \ l\ge 2,\ s\ge 0\ ,
  \label{eq:iter-estimate-2}
\end{equation}
and we set the following re-definitions
\begin{equation}
  \Fscr_r^{(r,\rgot;0)}\pluseq\Fscr_r^{(r,\rgot-1;\rgot)} \ ,
  \qquad
  \Fscr_r^{(r,\rgot;\rgot)}=0\ .
  \label{eq:iter-estimate-3}
\end{equation}
The new contributions to the perturbing terms are easily evaluated,
after having verified the following inequality:
\begin{equation}
  \left\|\frac{1}{j!}\lie{\Bgot^{(r,\rgot)}}^j f_l^{(r,\rgot-1;s)}\right\|
  \le \frac{\prod_{i=0}^{j-1}
  \left\{2\Big[(s+i\rgot)r+\big(l+i(r-1)\big)\rgot\Big]
  \Gscr^{(r,\rgot)}\right\}}{j!}\, \Fscr_l^{(r,\rgot-1;s)}\ ,
  \label{diseq:iter-estimate-4}
\end{equation}
that is easy to check by induction on $j$. Indeed, in the case $j=1$
one has simply to evaluate the coefficients generated by the
derivatives appearing in the Poisson brackets, while noticing that
$\lie{\Bgot^{(r,\rgot)}}^{j-1}
f_l^{(r,\rgot-1;s)}\in\Pscr_{l+(j-1)(r-1),2(s+(j-1)\rgot)}$ is essential
to deal with the generic case. Thus, the
prescriptions~\eqref{eq:f_l^rrgots_def_3}--\eqref{eq:f_l^rrgots_def_4}
can be translated into the following rules for the majorants:
\begin{equation}
\vcenter{\openup1\jot\halign{
 \hbox {\hfil $\displaystyle {#}$}
&\hbox {\hfil $\displaystyle {#}$\hfil}
&\hbox {$\displaystyle {#}$\hfil}\cr
\Fscr_{j(r-1)+1}^{(r,\rgot;j\rgot)} &\pluseq
&\frac{\prod_{i=1}^{j-1}
  \left\{2\Big[ir+\big(i(r-1)+1\big)\Big]\rgot
  \Gscr^{(r,\rgot)}\right\}}{j!}\, \Fscr_r^{(r,\rgot-1;\rgot)}
\qquad\ {\rm for}\ j\ge 2\ ,
\cr
\Fscr_{l+j(r-1)}^{(r,\rgot;s+j\rgot)} &\pluseq
&\frac{\prod_{i=0}^{j-1}
  \left\{2\Big[(s+i\rgot)r+\big(l+i(r-1)\big)\rgot\Big]
  \Gscr^{(r,\rgot)}\right\}}{j!}\, \Fscr_l^{(r,\rgot-1;s)}
\hfill\ {\rm for}\ l\ge 2\,,\ s\ge 0\,,\ j\ge 1\ .
\cr
}}
\label{eq:iter-estimate-5}
\end{equation}
This ends the description of the $(r,\rgot)$ iterative step
making part of our scheme of estimates.

Let us restart from our truncated expansions of
$\Hscr^{(2,0)}=H^{(\infty)}$, which have been determined as described
in subsect.~\ref{sbs:appl_KAM_2_sjsu_2D_sec}; the subsequent
computation of the majorants $\Fscr_l^{(r,\rgot;s)}$ is easy to code
and require a small amount of CPU time. In particular, we can provide
a simple rule to pass from the generating functions of a fixed degree
in the actions $\vet{p}$ to the next one. Actually, we define
$\Hscr^{(r+1,0)}=\Hscr^{(r,25(r-1))}$ and, correspondingly,
\begin{equation}
  \Fscr_l^{(r+1,0;s)}=\Fscr_l^{(r,25(r-1);s)} \ ,
  \qquad{\rm for}\ \ l\ge 2,\ s\ge 0\ .
  \label{eq:iter-estimate-6}
\end{equation}
This is due to the fact that $\Bgot^{(r,\rgot)}=0$
for $\rgot>25(r-1)$, because of the truncation up to
trigonometric degree $50$ of the computation of the Kolmogorov
normal form.

In particular, it is immediate to provide an estimate of the remainder
terms appearing in the Birkhoff normal form of degree $r$ in the
actions $\vet{p}$, that is
\begin{equation}
  \Hscr^{(r+1,0)}(\vet{p},\vet{q})= \vet{\omega}\cdot\vet{p}+
  \sum_{l=2}^{r}f_l^{(r+1,0;0)}(\vet{p})+
  \sum_{l>r}\Rscr_l^{(r+1)}(\vet{p},\vet{q})\ ,
\label{eq:Hscr^r+1,0}
\end{equation}
being $\Rscr_l^{(r+1)}=\sum_{s\ge 0} f_l^{(r+1,0;s)}$ for $l>r$.
In fact, the following inequality holds true:
\begin{equation}
  \|\Rscr_l^{(r+1)}\|\le\sum_{s=0}^{25(l-1)}\Fscr_l^{(r+1,0;s)}
  \quad{\rm for}\ l>r\ .
\label{diseq:iter-estimate-7}
\end{equation}

\subsection{Evaluation of the stability time for the secular model}\label{sbs:stab-time}

In this subsection, we adapt the approach developed
in~\cite{Gio-Loc-San-2009} to the present context. It is convenient to
consider the domain
\begin{equation}
\Delta_\rho = \left\{ \vet{p}\in\reali^n,\ |p_j|\leq\rho
\>,\>j=1\,,\,\ldots\,,\,n \right\}\ .
\label{frm6}
\end{equation}

\noindent
Therefore, for any generic function $g\in\Pscr_{l,2s}\,$,
the inequality
$$
\sup_{(\vet{p},\vet{q})\in\Delta_\rho\times\toro^n}
|g(\vet{p},\vet{q})| \le \|g\|\,\rho^l
$$
holds true as an immediate consequence of the
definition~\eqref{eq:def-norma} of $\|\cdot\|\,$.  Let us consider a
particular initial condition
$\big(\vet{p}(0),\vet{q}(0)\big)\in\Delta_{\rho_0}\times\toro^n$ with
$\rho_0<\rho$; let $T_e$ be the minimum value (for all
$\vet{q}(0)\in\toro^n$) for which the corresponding motion laws are such
that $\vet{p}(t)\in\Delta_{\rho}$ when $|t|<T_e$. Let us refer to
$T_e$ as the escape time from the domain $\Delta_\rho\,$.  This is the
quantity that we want to evaluate.  To this end we use the
elementary estimate
\begin{equation}
\left| p_j(t)-p_j(0) \right| \leq 
|t|\cdot\sup_{(\vet{p},\vet{q})\in\Delta_{\rho}\times\toro^n}|\dot p_j| <
|t|\,\sum_{l>r}\bigl\| \{ p_j,\Rscr_l^{(r+1)} \} \bigr\| {\rho^{l+1}}\ .
\label{frm7}
\end{equation}
The latter formula allows us to provide a lower estimate for the escape time
from the domain $\Delta_\rho\,$, namely
\begin{equation}
\tau(\rho_0,\rho,r) = 
\frac{\rho-\rho_0}
     {2\,\rho^{r+1}\,\sum_{s=0}^{25\,r}s\,\Fscr_{r+1}^{(r+1,0;s)}}
\ ,
\label{frm11}
\end{equation}
where we have kept just the first term of the series $\sum_{l>r}\| \{
p_j,\Rscr_l^{(r+1)} \} \rho^{l+1}\|$ because we restrict ourselves to
consider values of $\rho$ that are safely within its analyticity
domain. Moreover, the estimate of $\|\{ p_j,\Rscr_{r+1}^{(r+1)}\}\|$
is similar to~\eqref{diseq:iter-estimate-7} for $\Rscr_l^{(r+1)}$.

\begin{figure}
\centerline{\includegraphics[width=140mm]{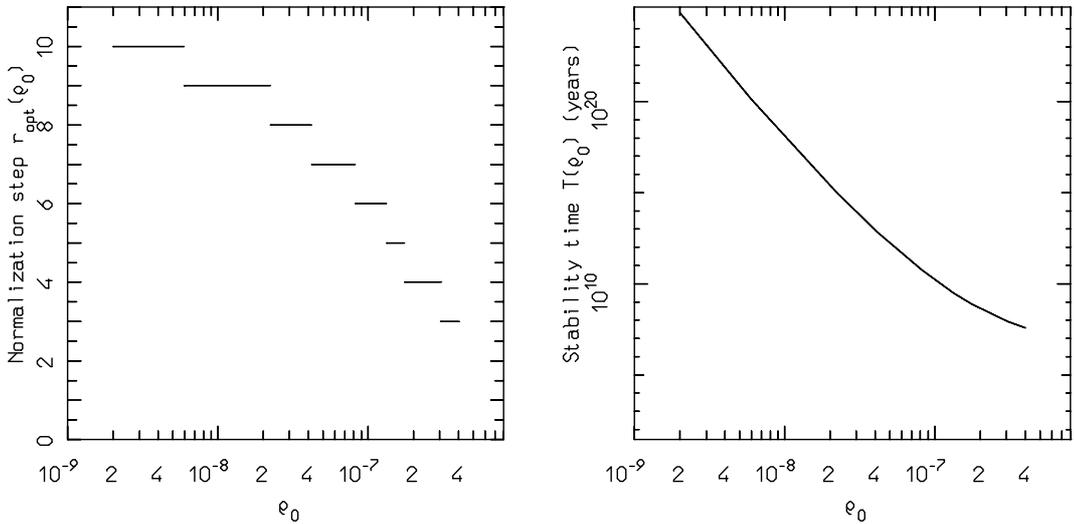}}
\caption{Optimal normalization order $r_{\rm opt}$ and estimated
  stability time $T(\rho_0)$ for our planar secular model of the
  system including Sun, Jupiter, Saturn and Uranus.  The time unit
  is the year.  See the text for more details.}
\label{fig:stab_time}
\end{figure}

In a practical application, usually $\rho_0$ is fixed
by the initial data, while $\rho$ and $r$ are left arbitrary.  Thus, we
proceed by looking for the maximal value of
$\tau(\rho_0,\rho,r)$ with respect to $\rho$ and $r$.
First we keep $r$ fixed and optimize the function
$\rho\mapsto\tau(\rho_0,\rho,r)$; this allows us to obtain
$\rho=\frac{r+1}{r} \rho_0$ and to introduce the
new function
\begin{equation}
\tilde\tau(\rho_0,r) = \sup_{\rho\ge\rho_0} \tau(\rho_0,\rho,r)
=\frac{r^{r}}{(r+1)^{r+1}}\,
\frac{1}{2\,\sum_{s=0}^{25\,r}\Big(s\,\Fscr_{r+1}^{(r+1,0;s)}\Big)\,\rho_0^{r}}
\ .
\label{eq:tildetau}
\end{equation}
Next we look for the optimal value $r_{\rm
opt}$ of $r$, which maximizes $\tilde\tau(\rho_0,r)$ when $r$ is
allowed to change.  This means that we look for the quantity
\begin{equation}
T(\rho_0) = \max_{r\ge 1} \tilde\tau(\rho_0,r)\ ,
\label{eq:T_di_rho0}
\end{equation}
which is our best estimate of the escape time, depending only on the
initial data.  We define the latter quantity as the estimated
stability time. After having compared the estimate
in~\eqref{diseq:iter-estimate-7} with the denominator appearing
in~\eqref{eq:tildetau}, it is natural to expect that the function
$r\mapsto\tilde\tau(\rho_0,r)$ asymptotically behaves in the same way
as the inverse of the remainder of the Birkhoff normal form, that is
$\tau(\rho_0,r)\sim C^r/[\rho^r(r!)^{n}]$, where $C$ is a suitable
constant and we assumed the optimal Diophantine condition for the
quasi-periodic motion on the initial KAM torus,
i.e. $|\vet{k}\cdot\vet{\omega}|>\gamma/|\vet{k}|^{n-1}$, for a fixed
value of $\gamma>0$. Let us recall that this kind of classical
estimates is essential to prove that the stability time grows
exponentially with respect to the inverse of the distance $\rho_0$
from that torus. On the other hand, for any fixed value of $\rho_0\,$,
$\tau(\rho_0,r)$ quickly reduces to zero, then there is a {\it finite}
optimal value $r_{\rm opt}\,$, maximizing $\tilde\tau(\rho_0,r)$.  In
the two plots of Figure~\ref{fig:stab_time}, we reported both the
step-wise function $r_{\rm opt}(\rho)$ and the corresponding estimated
stability time $T(\rho_0)=\tilde\tau(\rho_0,r_{\rm opt})$, according
to~\eqref{eq:T_di_rho0}. Actually, Figure~\ref{fig:stab_time} collects
the explicit results for our secular model. This has been possible,
because we preliminary computed the upper bounds
$\Fscr_{r+1}^{(r+1,0;s)}$ for $0\le s\le 25\,r\,$, $r\le
20\,$. In other words, starting from our truncated expansions of
$\Hscr^{(2,0)}=H^{(\infty)}$, we have explicitly calculated the
majorants for the first~19 normalization steps constructing the
Birkhoff normal form around the initial torus, whose corresponding
angular velocity vector is $\vet{\omega}$.

In our framework, the best proof for the effective stability of the
system would be produced by verifying that the set of possible initial
conditions (taking into account the limitations on the knowledge of
these values due to the observations) is covered by the domain
$\Delta_{\rho_0}\times\toro^n$.  In what follows, we make such a
comparison.

For what concerns the set of possible initial conditions, we can refer
to the values of $\Delta\xi_j$ and $\Delta\eta_j$ that are reported
for $j=1,\,2$ (corresponding to Jupiter and Saturn, respectively) in
Table~3 of~\cite{Gio-Loc-San-2009}: the uncertainties on the
determination of the secular actions are given by
\begin{equation}
\Delta I_j \simeq |\xi_j|\Delta\xi_j+|\eta_j|\Delta\eta_j\ ,
\label{eq:errori_assoluti_sulle_azioni}
\end{equation}
where the secular coordinates $\xi_j$ and $\eta_j$ can be calculated
by using formula~\eqref{var-Poincare-piano} and the initial conditions
listed in Table~\ref{tab:parameters_2D_SJSU}. Moreover,
in~\eqref{eq:errori_assoluti_sulle_azioni} we have assumed that the
effects induced by the diagonalizing canonical transformation
$(\vet{\xi},\vet{\eta})=\Dscr(\vet{x},\vet{y})$ are negligible; this
is actually confirmed by visual inspection of the coefficients
appearing in the matrix associated with the linear operator $\Dscr$.
All the subsequent changes of coordinates defined by the normalization
procedure are either near-to-the-identity or rigid translations; thus,
it is natural to assume that the values of the actions $\Delta p_j =
\Delta I_1+\Delta I_2\simeq 7\,\times 10^{-7}$,
for $j=1,\,2,\,3$, approximately gives us the radii of the ball
containing all the possible initial conditions, which are coherent
with the observations. Let us remark that the value of the secular
action of Uranus (that is about $22$ times lighter than Jupiter) is
one order of magnitude smaller than that of the major planet of our
Solar System; this explains why $\Delta I_3$ has not been involved in
the previous approximate evaluation of $\Delta p_j\,$.

\noindent
For what concerns our estimates, Figure~\ref{fig:stab_time} clearly
shows that our model is stable for a time comparable to the estimated age of the
Universe (around $10^{10}$ years) in a neighborhood of the initial torus
having a radius $\rho_0$ slightly smaller than $10^{-7}$.

\noindent
Therefore, our estimates about the domain of effective stability could
look slightly disappointing: its radius is just one order of magnitude
smaller with respect to that related to the possible initial
conditions of our model. Nevertheless, we think that our approach
could be improved so as to obtain a fully satisfying result. Indeed, the
evaluation of the ball radius of the initial conditions is quite
pessimistic, as it has been made by
following~\cite{Gio-Loc-San-2009} and, therefore, it is based on
observations made a few decades ago; the technological progress is
continuously reducing the uncertainties of the measures.  Moreover,
the radius $\rho_0$ of the effective stability domain can be
significantly enlarged, by performing explicit calculations for the
expansions of the Birkhoff normal form in the neighborhood of the
initial KAM torus. This can be preliminary done in such a way to
improve the results produced by the iterative scheme of
estimates\footnote{The effectiveness of an approach complementing
  explicit expansions of the Birkhoff normal form with rigorous
  computer-assisted estimates has been recently shown in the classical
  case-study concerning the Henon-Heiles system. These results are
  described in C. Caracciolo: ``Studio rigoroso della stabilit\`a
  effettiva di sistemi Hamiltoniani quasi-integrabili: stime
  computer-assisted'', thesis, Master in Mathematics, Univ. of Roma
  ``Tor Vergata'' (2016), that is available on request to the
  Author.}.

\subsection*{Acknowledgments}
This work has been supported by the research program ``Teorie
geometriche e analitiche dei sistemi Hamiltoniani in dimensioni finite
e infinite'', PRIN 2010JJ4KPA009, financed by MIUR.

\end{document}